\documentclass[sigconf, balance=false]{acmart}
\usepackage{popets}
\setcopyright{none}
\copyrightyear{2025}

\acmYear{}
\acmVolume{}
\acmNumber{}
\acmDOI{}
\acmConference{Proceedings on Privacy Enhancing Technologies}
\usepackage{algorithmic}
\usepackage[font=footnotesize]{subfig}
\usepackage{caption}
\usepackage{amsthm}
\usepackage{amsmath}
\usepackage{mathtools}
\usepackage{bbm}
\usepackage{stmaryrd}
\usepackage{booktabs}
\usepackage{tikz}
\usetikzlibrary{chains,positioning}
\allowdisplaybreaks

\newtheorem{proposition}{Proposition}
\newtheorem{theorem}{Theorem}

\newtheorem{remark}{Remark}
\theoremstyle{definition}
\newtheorem{definition}{Definition}

\newcommand{\concrete}{\texttt{Concrete}}

\usepackage{float}
\usepackage{makecell}
\usepackage[noend,lined,linesnumbered]{algorithm2e}

\usepackage{xcolor}
\definecolor{teal}{HTML}{008080}
\definecolor{lgtgray}{gray}{0.90}

\usepackage{hyperref}
\hypersetup{
   colorlinks=true,
   citecolor=teal
}

\usepackage{multirow}

\allowdisplaybreaks[1] 

\newcommand{\descr}[1]{\vspace{0.2cm} \noindent \textbf{#1}}

\newcommand{\revise}[1]{{\textcolor{black}{#1}}}

\begin{document}
%
\title{Practical, Private Assurance of the Value of Collaboration via Fully Homomorphic Encryption}{\thanks{This is the full version of the paper accepted for publication in the Proceedings on the 25th Privacy Enhancing Technologies Symposium (PoPETs) 2025.}}


\author{Hassan Jameel Asghar}
\affiliation{%
  \institution{Macquarie University}
  \country{Australia}}
\email{hassan.asghar@mq.edu.au}

\author{Zhigang Lu}
\affiliation{%
  \institution{Western Sydney University}
  \country{Australia}}
\email{z.lu@westernsydney.edu.au}

\author{Zhongrui Zhao}
\affiliation{%
  \institution{James Cook University}
  \country{Australia}}
\email{zhongrui.zhao@my.jcu.edu.au}

\author{Dali Kaafar}
\affiliation{%
  \institution{Macquarie University}
  \country{Australia}}
\email{dali.kaafar@mq.edu.au}



%

\begin{abstract}
Two parties wish to collaborate on their datasets. However, before they reveal their datasets to each other, the parties want to have the guarantee that the collaboration would be fruitful. We look at this problem from the point of view of machine learning, where one party is promised an improvement on its prediction model by incorporating data from the other party. The parties would only wish to collaborate further if the updated model shows an improvement in accuracy. Before this is ascertained, the two parties would not want to disclose their models and datasets. In this work, we construct an interactive protocol for this problem based on the fully homomorphic encryption scheme over the Torus (TFHE) and label differential privacy, where the underlying machine learning model is a neural network. Label differential privacy is used to ensure that computations are not done entirely in the encrypted domain, which is a significant bottleneck for neural network training according to the current state-of-the-art FHE implementations. We formally prove the security of our scheme assuming honest-but-curious parties, but where one party may not have any expertise in labelling its initial dataset. Experiments show that we can obtain the output, i.e., the accuracy of the updated model, with time many orders of magnitude faster than a protocol using entirely FHE operations. 
\end{abstract}


\maketitle


\section{Introduction}
Data collaboration, i.e., joining multiple datasets held by different parties, can be mutually beneficial to all parties involved as the joint dataset is likely to be more representative of the population than its constituents. In the real world, parties have little to no knowledge of each others' datasets before collaboration. Arguably, the parties would only collaborate if they had some level of trust in the quality of data held by other parties. When the parties involved are reputable organizations, one may assume their datasets to be of high quality. However, in many cases, little may be known about them. In such cases, each party would like some sort of \emph{assurance} that their collaboration will indeed be beneficial. 

Let us elaborate this scenario with an example. Assume companies $P_1$ and $P_2$ are in the business of developing antivirus products. Each company holds a dataset of malware programs labelled as a particular type of malware, e.g., ransomware, spyware and trojan. This labelling is done by a team of human experts employed by the company. Since manual labelling is expensive, $P_1$ uses a machine learning model to label new malware programs.
The performance of this model can be tested against a smaller \emph{holdout} dataset of the latest malware programs labelled by the same experts. However, due to a number of reasons such as the under-representation of some of the malware classes in the training dataset or \emph{concept drift}~\cite{widmer1996conceptdrift} between the training dataset and the holdout dataset, 
the performance of this model on the holdout dataset begs improvement. Company $P_2$ offers a solution: by combining $P_2$'s dataset with $P_1$'s, the resulting dataset would be more representative and hence would improve the accuracy of $P_1$'s model. Before going into the laborious process of a formal collaborative agreement with $P_2$, $P_1$ would like to know whether this claim will indeed be true. On the other hand, $P_2$ would not want to hand over its dataset, in particular, its labels to $P_1$ before the formal agreement. 

Many similar examples to the one outlined above may occur in real-world data collaboration scenarios. A rather straightforward solution to this problem can be obtained using fully homomorphic encryption (FHE). Party $P_2$ encrypts the labels of its dataset using the encryption function of the FHE scheme,\footnote{For reasons discussed in Section~\ref{subsec:setting}, we only consider the case when the labels are encrypted, and not the features themselves.} and sends its dataset with encrypted labels to $P_1$. Party $P_1$, then combines this dataset with its own, trains the model and tests its accuracy against the holdout dataset all using FHE operations. The final output is then decrypted by party $P_2$.\footnote{We are assuming that decryption is faithful.} However, this involves training the model entirely in the encrypted domain, which even with the state-of-the-art homomorphic encryption schemes, is computationally expensive~\cite{hesamifard2018privacy,nandakumar2019towards,xu2019cryptonn,lou2020glyph,xu2021nn}. 

In this paper, we propose an efficient solution to this problem. \revise{Our main contributions are as follows
\begin{itemize}
    \item We combine fully homomorphic encryption over the torus~\cite{chillotti2020tfhe} with label differential privacy~\cite{label-dp} to provide an interactive solution to the problem where parties $P_1$ and $P_2$ would like to train a model on their combined datasets to assess its accuracy without learning the resulting model and each other's data labels. Our main idea is that since the features are known in the clear, the first forward pass in the backpropagation algorithm of neural networks can be performed in the clear, up to the point where we utilize the (encrypted) labels from party $P_2$'s dataset. If further computation is done homomorphically, then we would endure the same computational performance bottleneck as previous work. We, therefore, add (label) differentially private noise to the gradients and decrypt them before the backward pass. This ensures that most steps in the neural network training are done in cleartext, albeit with differentially private noise, giving us computational performance improvements over an end-to-end FHE solution. 
    \item We use label differential privacy in a novel way. More specifically, $P_2$ can use a specific value of the differentially privacy parameter $\epsilon$ such that the accuracy of the model on the joint datasets lies between the accuracy of $P_1$'s model and the accuracy achievable on the joint model if no differential privacy were to be applied. This provides assurance to $P_1$ that the combined dataset will improve accuracy promising further improvement if the parties combine their datasets in the clear via the formal collaborative agreement. Thus, in our protocol, we need not worry about making the differentially private model as accurate as the final model, which is in general a difficult task in differential privacy literature. 
    \item We initiate the study to determine whether it is possible to improve model accuracy by joining two datasets where the labels from the second dataset are not labelled with any domain knowledge. As we show, there are cases in which the combined dataset may show improvement over the standalone dataset even without domain knowledge. This problem has broader applications than our work. Our treatment is limited to binary classification and the 0-1 loss function. We leave full exploration of this topic as future work. 
    \item We provide an implementation of our protocol using Zama's \concrete{} library and evaluate our protocol over multiple datasets. Our results show that our protocol can be implemented faster than an end-to-end FHE solution. 
\end{itemize}
}

\section{Preliminaries and Threat Model}
\label{sec:back}

\subsection{Notation}

We follow the notations introduced in~\cite{shai-shai-book}. The datasets come from the joint domain: $\mathbb{D} = \mathcal{X} \times \mathcal{Y}$, where $\mathcal{X}$ denotes the domain of features, and $\mathcal{Y}$ the domain of labels. A dataset $D$ is a multiset of elements drawn i.i.d. under the joint distribution $\mathcal{D}$ over domain points and labels. We denote by $\mathcal{D}_x$, the marginal distribution of unlabelled domain points. In some cases, a dataset may be constructed by drawing unlabelled domain points under $\mathcal{D}_x$, and then labelled according to some labelling function, which may not follow the marginal distribution of labels under $\mathcal{D}$, denoted $\mathcal{D}_{y|x}$. In this case, we shall say that the dataset is labelled by the labelling function to distinguish it from typical datasets. Let $\mathcal{A}$ denote the learning algorithm, e.g., a neural network training algorithm. We denote by $M \leftarrow \mathcal{A}(D)$ the model $M$ returned by the learning algorithm on dataset $D$. 
\revise{The notation $(x, y) \sim \mathcal{D}$ means that the sample $(x, y)$ is sampled from the distribution $\mathcal{D}$, and the notation $(x, y) \sim D$, where $D$ is a dataset means that the sample $(x, y)$ is sampled uniformly at random from $D$.} Given the model $M$, and a sample $(x, y) \sim \mathcal{D}$, we define a generic loss function $\ell(M, x, y)$, which outputs a non-negative real number. For instance, $\ell(M, x, y)$ can be the 0-1 loss function, defined as: 
$\ell(M, x, y) = 1$ if $M(x) \neq y$, and 0 otherwise. We define the true error of $M$ as $
L_{\mathcal{D}}(M) = \mathbb{E}_{(x, y) \sim \mathcal{D}}[\ell(M, x, y)]$. Notice that for the 0-1 loss function, this means that
$L_{\mathcal{D}}(M) = \Pr_{(x, y) \sim \mathcal{D}} [M(x) \neq y]$. The empirical error of the model $M$ over the dataset $D$ having $m$ elements $(x_1, y_1), \ldots, (x_m, y_m)$ is defined as:
\begin{equation}
\label{eq:emp-loss}
L_{D}(M) = \frac{1}{m} \sum_{i \in [m]} \ell(M, x_i, y_i)
\end{equation}
\revise{Note that for the $0-1$ loss function $L_D(M) = \Pr_{(x, y) \sim D} [M(x) \neq y]$.
}

\subsection{The Setting}
\label{subsec:setting}
\descr{The Scenario.} We consider two parties $P_1$ and $P_2$. For $i \in \{1, 2\}$, party $P_i$'s dataset is denoted $D_i$. 
Each $D_i$ contains points of the form $(\mathbf{x}, y)$. 
The parties wish to collaborate on their datasets $D_{i}$. 
The features $\mathbf{x}$ are shared in the open; whereas the labels $y$ for each $\mathbf{x}$ in $D_i$ are to be kept secret from the other party. This scenario holds in applications where \textit{gathering data ($\mathbf{x}$) may be easy, but labelling is expensive}. For example, malware datasets (binaries of malware programs) are generally available to antivirus vendors, and often times features are extracted from these binaries using publicly known feature extraction techniques, such as the LIEF project~\cite{lief}. However, labelling them with appropriate labels requires considerable work from (human) experts. Other examples include sentiment analysis on public social media posts, and situations where demographic information is public (e.g., census data) but income predictability is private~\cite{label-dp}. \revise{Lastly, we want to acknowledge that our focus is on supervised learning. There are works that show unsupervised learning techniques may improve the model~\cite{zahoora2022ransomware} even if the labels are important in many applications such as malware detection. Our focus is on the added value of the labels.}

\descr{The Model.} Before the two parties reveal their datasets to each other, the parties want to have the guarantee that the collaboration would be \textit{valuable}. We shall assume that party $P_1$ already has a model $M_1$ trained on data $D_1$. $P_1$ also has a labelled holdout data $D_\text{hold}$ against which $P_1$ tests the accuracy of $M_1$. The goal of the interaction is to obtain a new model $M_2$ trained on $D_1 \cup D_2$. $M_{2}$'s accuracy again is tested against $D_\text{hold}$. The collaboration is defined to be \textit{valuable} for $P_{1}$ if the accuracy of $M_2$ is higher than the accuracy of $M_1$ against $D_\text{hold}$. In this paper, we only study  \textit{value} from $P_{1}$'s perspective. We will consider a neural network trained via stochastic gradient descent as our canonical model.

\descr{The Holdout Dataset.} As mentioned above, $P_1$ has a holdout dataset $D_\text{hold}$ against which the accuracy of the models is evaluated. This is kept separate from the usual training-testing split of the dataset $D_1$. It makes sense to keep the same holdout dataset to check how the model trained on the augmented/collaborated data performs. For instance, in many machine learning competitions teams compete by training their machine learning models on a publicly available training dataset, but the final ranking of the models, known as \emph{private leaderboard}, is done on a hidden test dataset~\cite{blum2015ladder}. This ensures that the models are not overfitted by using the test dataset as feedback for re-training. We assume that $D_\text{hold}$ is continually updated by adding new samples (e.g., malware never seen by $P_{1}$) labelled by human experts, and is more representative of the population than $D_1$. For instance, the holdout dataset reflects the \textit{concept drift}~\cite{widmer1996conceptdrift} better than $D_{1}$. If $D_{2}$ happens to have the same \textit{concept drift} as $D_{\text{hold}}$, then $M_{2}$ (trained on $D_1 \cup D_2$) would have better test accuracy than $M_{1}$ against $D_{\text{hold}}$. Alternatively, the holdout dataset could be more balanced than $D_{1}$, e.g., if $D_{1}$ has labels heavily skewed towards one class. Again, in this case, if $D_2$ is more balanced than $D_1$, then $M_2$ will show better accuracy. We argue that it is easy for $P_1$ to update $D_\text{hold}$ than $D_1$ as the latter requires more resources due to the difference in size. 

\subsection{Privacy}

\descr{Privacy Expectations.} We target the following privacy properties:
\begin{itemize}
    \item Datasets $D_1$ and $D_\text{hold}$, and model $M_1$ should be hidden from $P_2$.
    \item The labels of dataset $D_2$ should be hidden from $P_1$.
    \item Neither $P_1$ nor $P_2$ should learn $M_2$, i.e., the model trained on $D_1 \cup D_2$.
    \item Both parties should learn whether ${L}_{\text{hold}}(M_2) < {L}_{\text{hold}}(M_1)$, where ${L}_{\text{hold}}$ is the loss evaluated on $D_\text{hold}$ (see Eq.~\eqref{eq:emp-loss}).
\end{itemize}

\descr{Threat Model.} We assume that the parties involved, $P_1$ and $P_2$, are honest-but-curious. This is a reasonable assumption since once collaboration is agreed upon, the model trained on clear data should be able to reproduce any tests to assess the quality of data pre-agreement. Why then would $P_1$ not trust the labelling from $P_2$? This could be due to the low quality of $P_{2}$'s labels, for many reasons. For example, $P_2$'s expertise could in reality be below par. In this case, even though the labelling is done honestly, it may not be of sufficient quality. Furthermore, $P_2$ can in fact lie about faithfully doing its labelling without the fear of being caught. This is due to the fact that technically there is no means available to $P_1$ to assess how $P_{2}$'s labels were produced.
All $P_2$ needs to do is to provide the same labels before and after the collaborative agreement. As long as labelling is consistent, there is no fear of being caught.

\subsection{Background}
\label{subsec:bg}

\descr{Feedforward Neural Networks.}
A fully connected feedforward neural network is modelled as a graph with a set of vertices (neurons) organised into layers and weighted edges connecting vertices in adjacent layers. The sets of vertices from each layer form a disjoint set. There are at least three layers in a neural network, one input layer, one or more hidden layers and one output layer. 
The number of neurons in the input layer equals the number of features (dimensions). 
The number of neurons in the output layer is equal to the number of classes $K$. The vector of weights $\mathbf{w}$ of all the weights of the edges constitutes the parameters of the network, to be learnt during training. We let $R$ denote the number of weights in the last layer, i.e., the number of edges connecting to the neurons in the output layer. For a more detailed description of neural networks, see~\cite{shai-shai-book}.   


\descr{Backward Propagation and Loss.} Training a deep neural network involves multiple iterations/epochs of forward and backward propagation. The input to each neuron is the weighted sum of the outputs of all neurons connected to it (from the previous layer), where the weights are associated with the edge connecting to the current neuron.
The output of the neuron is the application of the activation function on this input. In this paper, we assume the activation function to be the sigmoid function. Let $\mathbf{z} = \begin{pmatrix} z_1, z_2, \ldots, z_K\end{pmatrix}$ denote the output of the last layer of the neural network. We assume there to be a softmax layer, immediately succeeding this which outputs the probability vector $\mathbf{p}$, whose individual components are given as $p_i = {e^{z_i}}/(\sum_j e^{z_j})$.
Clearly, the sum of these probabilities is 1. Given the one-hot encoded label $\mathbf{y}$, one can compute the loss as $L(\mathbf{w}) = \sum_{i = 1}^K \ell(p_i, y_i)$, where $\ell(\cdot, \cdot)$ is the loss function. In this paper, we shall consider it to be the cross-entropy loss given by: $\ell(p_i, y_i) = - \sum_{i = 1}^K y_i \ln p_i$. Given $L(\mathbf{w})$, we can calculate its gradient as:
\begin{equation*}
    \nabla L(\mathbf{w}) = \frac{\partial L}{\partial \mathbf{w}} = \frac{\partial L}{\partial \mathbf{p}} \frac{\partial \mathbf{p}}{\partial \mathbf{z}} \frac{\partial \mathbf{z}}{\partial \mathbf{w}} 
\end{equation*}
This chain rule can be used in the backpropagation algorithm to calculate the gradients to minimise the loss function via stochastic gradient descent (SGD). 
The calculated gradients will be then used to update the weights associated with each edge for the forward propagation in the next epoch.
For more details of the backpropagation-based SGD algorithm, see~\cite{shai-shai-book}.

\descr{Label Differential Privacy.} Since only the labels of the dataset $D_2$ are needed to be private, we shall use the notion of label differential privacy to protect them. Ordinary differential privacy~\cite{dwork2006calibrating} defines neighbouring datasets as a pair of datasets which differ in one row. Label differential privacy considers two datasets of the same length as being neighbours if they differ only in the \emph{label} of exactly one row~\cite{label-dp}. This is a suitable definition of privacy for many machine learning applications such as malware labels (as already discussed) and datasets where demographic information is already public but some sensitive feature (e.g., income) needs to be protected~\cite{label-dp}. Furthermore, for tighter privacy budget analysis we use the notion of $f$-differential privacy~\cite{dong2022gaussian}, modified for label privacy since it allows straightforward composition of the Gaussian mechanism as opposed to the normal definition of differential privacy which needs to invoke its approximate variant to handle this mechanism. 
More formally, two datasets $D_1$ and $D_2$ are said to be neighbouring datasets, if they are of the same size and differ only in at most one label. 

The $f$-differential privacy framework is based on the hypothesis testing interpretation of differential privacy. Given the output of the mechanism $\mathcal{A}$, the goal is to distinguish between two competing hypotheses: the underlying data set being $D_1$ or $D_2$. Let $Q_1$ and $Q_2$ denote the probability distributions of $\mathcal{A}(D_1)$ and $\mathcal{A}(D_2)$, respectively. Given any rejection rule $0 \le \phi \le 1$, the type-I and type-II errors are defined as follows ~\cite{dong2022gaussian}: $\alpha_\phi = \mathbb{E}_{Q_1}[\phi]$ and  $\beta_\phi = 1-  \mathbb{E}_{Q_2}[\phi]$.

\begin{definition}[Trade-off function~\cite{dong2022gaussian}]
\label{def:T}
For any two probability distributions $Q_1$ and $Q_2$ on the same
space, the \emph{trade-off function} $T(Q_1, Q_2) : [0, 1] \to [0, 1]$ is defined by
\[
T(Q_1, Q_2)(\alpha) = \inf \{\beta_\phi : \alpha_\phi \leq \alpha\}
\]

\noindent for all $\alpha \in [0, 1]$, where the infimum is taken over all (measurable) rejection rules.
\end{definition}

A trade-off function gives the minimum achievable type-II error at any given level of type-I error. For a function to be a trade-off function, it must satisfy the following conditions.

\begin{proposition}[\cite{dong2022gaussian}]
A function $f : [0, 1] \to [0, 1]$ is a trade-off function if and only if $f$ is convex, continuous and non-increasing, and $f(x) \leq 1 - x$ for all $x \in [0, 1]$.
\end{proposition}

Abusing notation, let $\mathcal{A}(D)$ denote the distribution of a mechanism $\mathcal{A}$ when given a data set $D$ as input. We now give a definition of $f$-label differential privacy based on the definition of ordinary $f$-differential privacy:
\begin{definition}[$f$-Label Differential Privacy~\cite{dong2022gaussian}]
Let $f$ be a trade-off function. A mechanism $\mathcal{A}$ is said to
be \emph{$f$-label differentially private} if $T(\mathcal{A}(D_1), \mathcal{A}(D_2)) \geq f$ for all neighbouring data sets $D_1$ and $D_2$.
\end{definition}
Note that the only change here from the definition of $f$-differential privacy from~\cite{dong2022gaussian} is how we define neighbouring datasets. In our case neighbouring datasets differ only in the label of at most one sample. We now give a concrete $f$-label differentially private mechanism.

\begin{proposition}[$\epsilon$-Gaussian Label Differential Privacy~\cite{dong2022gaussian}]
\label{prop:gldp}
The mechanism $q(D) + \mathcal{N}(0, \Delta q^2/\epsilon^2)$  is $\epsilon$-GLDP where $\Delta q$ is the sensitivity of the function $q$ over any pairs of neighbouring (in label) datasets $D_1$ and $D_2$.
\end{proposition}

The notion of $\epsilon$-GLDP satisfies both sequential and parallel composition.

\begin{proposition}[Sequential and Parallel Composition]
\label{prop:seq-par-comp}
The composition of $n$-fold (sequential) $\epsilon_i$-GLDP mechanisms is $\sqrt{\epsilon_1^2 + \cdots + \epsilon_n^2}$-GLDP~\cite{dong2022gaussian}. Let a sequence of $n$ mechanisms $\mathcal{A}_i$ each be $\epsilon_i$-GLDP. Let $\mathbb{D}_i$ be disjoint subsets of the data domain $\mathbb{D}$. The joint mechanism defined as the sequence of $\mathcal{A}_i(D \cap \mathbb{D}_i)$ (given also the output of the previous $i-1$ mechanisms) is $\max\{\epsilon_1, \ldots, \epsilon_n\}$-GLDP~\cite{smith2022making}.
\end{proposition}

Lastly, $f$-differential privacy is also immune to post-processing~\cite{dong2022gaussian}. That is, applying a randomised map with an arbitrary range to the output of an $f$-label differentially private algorithm maintains $f$-label differential privacy.

\descr{Fully Homomorphic Encryption over the Torus (TFHE).} Let $\mathbb{T}$ denote the torus, the set of real numbers modulo 1, i.e., the set $[0, 1)$. The torus defines an Abelian group where two elements can be added modulo 1. The internal product is not defined. However, one can multiply an integer with a torus element by simply multiplying them in a usual way and reducing modulo 1. For a positive integer $q$, the discretised torus $\mathbb{T}_q$ is defined as the set $\{0, \frac{1}{q}, \ldots, \frac{q-1}{q}\}$. Clearly $\mathbb{T}_q \subset \mathbb{T}$. Given elements $\frac{a}{q}, \frac{b}{q} \in \mathbb{T}_q$, their sum is $\frac{c}{q} \in \mathbb{T}_q$, where $c \equiv a + b \pmod{q}$. Given an integer $z \in \mathbb{Z}$ and a torus element $\frac{a}{q} \in \mathbb{T}_q$, we define their product $z \cdot \frac{a}{q}$ as $\frac{b}{q}$ where $b \equiv za \pmod{q}$. The plaintext space is a subgroup of $\mathbb{T}_q$, defined as $\mathcal{P} = \{0, \frac{1}{p}, \ldots, \frac{p-1}{p}\}$, for some $p \ge 2$ such that $p$ divides $q$. 

Let $\chi$ be the normal distribution $\mathcal{N}(0, \sigma^2)$ over $\mathbb{R}$. Let $e_0 \leftarrow \chi$. The noise error $e \in \hat{\chi}$ is defined as $e = \frac{\lfloor q e_0 \rceil}{q}$.
Let $N$ be a positive integer. Let $\mathbf{s} = (s_1, \ldots, s_N)$ be a binary vector chosen uniformly at random (the private key). Given a message $x \in \mathcal{P}$, the TLWE encryption of $x$ under $\mathbf{s}$ is defined as $\mathbf{c} = (\mathbf{a}, b) \in \mathbb{T}_{q}^{N+1}$, where $\mathbf{a}$ is a vector of $N$-elements drawn uniformly at random from $\mathbb{T}_q$, and
\[
b = \langle \mathbf{s}, \mathbf{a} \rangle + x + e.
\]
To decrypt $\mathbf{c}$ one computes: $
x^* = b - \langle \mathbf{s}, \mathbf{a} \rangle$, 
and returns the nearest element in $\mathcal{P}$ to $x^*$. The scheme is secure under the learning with errors (LWE) problem over the discretized torus~\cite{chillotti2020tfhe, joye2021guide}:

\begin{definition}[TLWE Assumption]
\label{def:tlwe-assumption}
Let $q, N \in \mathbf{N}$. Let $\mathbf{s} = (s_1, \ldots, s_N)$ be a binary vector chosen uniformly at random. Let $\hat{\chi}$ be an error distribution defined above. The learning with errors over the discretized torus (TLWE) problem is to distinguish samples chosen according to the following distributions:
\[
\mathcal{D}_0 = \{ (\mathbf{a}, b) \mid \mathbf{a} \leftarrow \mathbb{T}_q^N, b \leftarrow \mathbb{T}_q \},
\]
and
\[
\mathcal{D}_1 = \{ (\mathbf{a}, b) \mid \mathbf{a} \leftarrow \mathbb{T}_q^N, b = \langle \mathbf{s}, \mathbf{a} \rangle + e, e \leftarrow \hat{\chi}\},
\]
where except for $e$ which is sample according to the distribution $\hat{\chi}$, the rest are sampled uniformly at random from the respective sets. 
\end{definition}

\descr{Integer Encoding.} Before encrypting, we will encode the input as an integer. This is a requirement in the current version of Zama's \concrete{} TFHE library~\cite{concrete}, which we use for our implementation. Let $\mathbf{x} \in \mathbb{R}^m$. For a precision level $r$, where $r$ is a positive integer, we will encode $\mathbf{x}$ as $\lfloor r \mathbf{x} \rfloor = (\lfloor r x_1 \rfloor, \ldots ,\lfloor r x_m \rfloor)$, before encrypting. Decoding is done, after decryption, by dividing the encoded vector by $r$. For instance if $r = 10$, then the real number $3.456$ is encoded as $\lfloor 34.56 \rfloor = 34$. Decoding it yields $3.4$. 

\descr{Parameters.} We use the default parameters for TLWE encryption available through the \concrete{} library. Under the default setting, we have $N = 630$, giving us a security level of 128 bits. The parameter $q$ is set to 64 bits, meaning a torus element can be represented by 64 bits~\cite{joye2021guide}. Since the plaintext parameter $p$ divides $q$, we can assume $p$ to be less than 64 bits. The noise parameter $\sigma$ is $2^{-15}$~\cite{joye2021guide}.




\section{Lack of Domain Knowledge}
\label{sec:accuracy-gain}
Before we give a privacy-preserving solution to our problem, we want to investigate whether the problem has a solution in the clear domain. More precisely, we seek to find conditions when party $P_2$, lacking domain knowledge, cannot come up with a dataset $D_2$ such that ${L}_{\text{hold}}(M_2) < {L}_{\text{hold}}(M_1)$. Here $M_2 \leftarrow \mathcal{A}(D_1 \cup D_2)$ and $M_1 \leftarrow \mathcal{A}(D_1)$, and $L_{\text{hold}}$ is as defined in Eq.~\eqref{eq:emp-loss}. Although at first glance the problem appears to be simple, in reality it is quite involved as multiple scenarios need to be considered. We therefore only assume binary classification, with the results for the multiclass setting left for future work. We assume the 0-1 loss function. 
How do we define lack of domain knowledge? Since the feature vectors are public, $P_2$ can easily obtain a set of raw inputs to obtain the feature vectors in $D_2$. Thus, domain knowledge should be captured in the labels to the feature vectors in $D_2$. We define lack of domain knowledge as
using a labeling function $g$, \revise{potentially probabilistic}, which labels any domain point $x$ independent of the distribution of its true label. That is given any $x \in \mathcal{X}$:
\begin{equation}
\label{eq:lab-func}
\Pr[g(x) = y' \mid \mathcal{D}_{y|x}] = \Pr[g(x) = y' ],
\end{equation}
for all $y' \in \mathcal{Y}$. Note that this does not mean that the labeling is necessarily incorrect. We call $g$ as defined in Eq.~\ref{eq:lab-func} as an \emph{oblivious labeling function}. 
We shall first show that if $D$ is labelled by an oblivious labeling function $g$, then $M$ as returned by a learning algorithm $\mathcal{A}$ (taking $D$ as the input) will have:
\revise{
 \[
\mathbb{E}_g[{L}_{\text{hold}}(M)] = 1/2,
 \]
 provided that $D_\text{hold}$ is a \emph{balanced dataset}, where the subscript means that expectation is taken over the random choices of $g$.} Having established the necessity of $D_\text{hold}$ to be balanced, we shall then show that
\[
\mathbb{E}_g[{L}_{\text{hold}}(M_2)] \geq {L}_{\text{hold}}(M_1),
\]
where $M_2 \leftarrow \mathcal{A}(D_1 \cup D_2)$, and $M_1 \leftarrow \mathcal{A}(D_1)$, provided a certain condition is met. Details follow.

\begin{theorem}
\label{lem:half-loss-balanced}
Let $D_\text{hold}$ be a balanced dataset. That is
$\Pr[y = 1 \mid (x, y) \sim D_\text{hold}] = \frac{1}{2}$, where $\sim$ denotes uniform random sampling. Let $D$ be a dataset labelled by an oblivious labeling function $g$. Let $\mathcal{A}$ be a learning algorithm. Let $M \leftarrow \mathcal{A}(D)$. \revise{Then, 
\[
\mathbb{E}_g[{L}_{\text{hold}}(M)] = 1/2,
\]
where the expectation is over the random choices of $g$.}
\end{theorem}
\begin{proof}
See Appendix~\ref{app:knowledge-proofs}
\end{proof}

What happens if $D_\text{hold}$ is not balanced? Then, we may get a loss less than $1/2$. To see this, assume that $\Pr[y = 1 \mid (x, y) \sim D_\text{hold}] = q > \frac{1}{2}$. Consider the oblivious labelling function $g$ which outputs 1 with probability $q = 1$, i.e., the constant function $g(x) = 1$. Then if $M$ is the Bayes optimal classifier~\cite[\S 3.2.1]{shai-shai-book} for $g$, then for $(x, y) \sim D_\text{hold}$, we have:
\begin{align*}
    \Pr [M(x) \neq y] &= \Pr [M(x) = 0 \mid y = 1 ] \Pr [y = 1] \\
    &+ \Pr [M(x) = 1 \mid y = 0] \Pr [y = 0]\\
    &= \Pr [M(x) = 1 \mid y = 0] \Pr [y = 0]\\
    &=  \Pr [y = 0] = 1 - q < \frac{1}{2}.
\end{align*}
Thus, it is crucial to test the model over a balanced dataset. 
\begin{theorem}
\label{theo:m2-less-than-m1}
Let $D_1$ be a dataset. Let $D_2$ be a dataset labelled according to an oblivious labeling function $g$. Let $D_\text{hold}$ be a balanced dataset. Let $\mathcal{A}$ be a learning algorithm. Let $M_1 \leftarrow \mathcal{A}(D_1)$ and $M_2 \leftarrow \mathcal{A}(D_1 \cup D_2)$. \revise{If :
\begin{align}
 &\Pr_{g, (x,y) \sim D_\text{hold}} [M_2(x) = 0 \mid y = 1 ] + \Pr_{g, (x,y) \sim D_\text{hold}} [M_2(x) = 1 \mid y = 0] \nonumber\\
 &\geq \Pr_{(x,y) \sim D_\text{hold}} [M_1(x) = 0 \mid y = 1 ] + \Pr_{ (x,y) \sim D_\text{hold}} [M_1(x) = 1 \mid y = 0] \label{eq:oea},
\end{align}
then
\[
\mathbb{E}_g[{L}_{\text{hold}}(M_2)] \geq {L}_{\text{hold}}(M_1),
\]
where the expectation is taken only over random choices of $g$, and $(x, y) \sim D_\text{hold}$ means the sample is chosen uniformly at random from $D_\text{hold}$.}
\end{theorem}
\begin{proof}
See Appendix~\ref{app:knowledge-proofs}.
\end{proof}

\revise{Concluding, we have:
\begin{theorem}
 \label{theo:hoeffding}
Let $D_1$ be a dataset. Let $D_2$ be a dataset labelled according to an oblivious labeling function $g$. Let $D_\text{hold}$ be a balanced dataset of size $m$. Let $\mathcal{A}$ be a learning algorithm. Let $M_1 \leftarrow \mathcal{A}(D_1)$ and $M_2 \leftarrow \mathcal{A}(D_1 \cup D_2)$. Under condition~\eqref{eq:oea} of Theorem~\ref{theo:m2-less-than-m1}, for any $\delta > 0$
\[
\Pr_g[L_\text{hold}(M_1) - L_\text{hold}(M_2) \geq \delta] \leq \exp(-2m\delta^2),
\]
where the probability is taken over the random choices of $g$.   
\end{theorem}
\begin{proof}
See Appendix~\ref{app:knowledge-proofs}.
\end{proof}
}

\revise{Exactly when does condition \eqref{eq:oea} in Theorems~\ref{theo:m2-less-than-m1} and \ref{theo:hoeffding} hold? We postulate and give some experimental evidence that the condition holds only when $D_1$ follows the distribution $\mathcal{D}$. On the other hand, if the labels of $D_1$ are out of distribution, i.e., one class is more under-represented compared to the true distribution, the above condition does not hold. Due to lack of space, details appear in Appendix~\ref{app:knowledge-proofs}. We acknowledge that we have only touched the tip of the iceberg when it comes to the full exploration of this problem for both binary and multi-class setting and leave it as an open problem.}
 
\section{Our Solution}

\subsection{Intuition}
\label{sec:intuition}
\begin{figure}
\centering
\begin{tikzpicture}[scale=0.65]
\draw[lightgray, fill, very thick, rounded corners] (0,0) rectangle (3,1);
\draw[lightgray, fill, very thick, rounded corners] (0,1.1) rectangle (3,3.1);
\draw[black, fill, very thick, rounded corners] (3.1,0) rectangle (3.6,1);
\draw[lightgray, fill, very thick, rounded corners] (3.1,1.1) rectangle (3.6,3.1);

\node at (-0.5, 0.5) {$D_2$};
\node at (-0.5, 2) {$D_1$};

\node at (1.5, 3.4) {${x}$};
\node at (3.4, 3.4) {$y$};

\end{tikzpicture}
\caption{A batch will contain some points from $D_1$ and some from $D_2$. Only the labels from the points in $D_2$ needs to be kept private (shaded black) from $P_1$.}
\label{fig:batch}
\end{figure}

Consider the training of the neural network on $D_1 \cup D_2$, with weights $\mathbf{w}$. Using the stochastic gradient descent (SGD) algorithm, one samples a batch $B$, from which we calculate per sample loss $L_s(\mathbf{w})$, where $s = (\mathbf{x}, \mathbf{y}) \in B$. Given this, we can compute the average loss over the batch via:
\begin{equation}
\label{eq:gradients}
L_B(\mathbf{w}) = \frac{1}{|B|} \sum_{s \in B} L_s(\mathbf{w}) = \frac{1}{|B|} \sum_{\substack{s \in B \\ s \in D_1}} L_s(\mathbf{w}) + \frac{1}{|B|} \sum_{\substack{s \in B\\ s \in D_2}} L_s(\mathbf{w}) 
\end{equation}
As shown in Figure~\ref{fig:batch}, everything in this computation is known to $P_1$, except for the labels in $D_2$. Thus, $P_1$ can compute the gradients for samples in $D_1$, but to update the weights, $P_1$ needs the gradients for samples from $D_2$. From Eq.~\eqref{eq:gradients}, we are interested in computing the loss through the samples in a batch $B$ that belong to the dataset $D_2$. Overloading notation, we still use $B$ to denote the samples belonging to $D_2$. The algorithm to minimize the loss is the stochastic gradient descent algorithm using backpropagation. This inolves calculating the gradient $\nabla L_B(\mathbf{w})$. As noted in~\cite{yuan2021label}, if we are using the backpropagation algorithm, we only need to be concerned about the gradients corresponding to the last layer. Again, to simplify notation, we denote the vector of weights in the last layer by $\mathbf{w}$. 

In Appendix~\ref{app:grad}, we show that the gradient of the loss can be computed as:
\begin{equation}
    \nabla L_B(\mathbf{w})  = \frac{1}{|B|} \sum_{s \in B} \sum_{i = 1}^K p_i(s) \frac{\partial z_i(s)}{\partial \mathbf{w}} - \frac{1}{|B|} \sum_{s \in B} \sum_{i = 1}^K y_i(s) \frac{\partial z_i(s)}{\partial \mathbf{w}}, \label{eq:two-terms}
\end{equation}
where $y_i(s)$ is the $i$th label of the sample $s$, $p_i(s)$ is the probability of the $i$th label of sample $s$, $z_i(s)$ is the $i$th input to the softmax function for the sample $s$, and $K$ denotes the number of classes. The LHS term of this equation can be computed by $P_1$ as this is in the clear. However, the RHS term requires access to the labels. 

If we encrypt the labels, the gradients calculated in Eq.~\eqref{eq:two-terms} will be encrypted, using the homomorphic property of the encryption scheme. This means that the gradient of the batch, as well as the weight updates will be encrypted. Thus, the entire training process after the first forward pass of the first epoch will be in the encrypted domain. While this presents one solution to our problem, i.e., obtaining an encrypted trained model, which could then be decrypted once the two parties wish to collaborate, existing line of works~\cite{hesamifard2018privacy,nandakumar2019towards,xu2019cryptonn,lou2020glyph,xu2021nn} show that neural network training entirely in the encrypted domain is highly inefficient. For instance, a single mini-batch of 60 samples can take anywhere from more than 30 seconds to several days with dedicated memory ranging from 16GB to 250GB using functional encryption or homomorphic encryption~\cite{xu2021nn}. In some of our neural network implementations, we use a batch size of 128 with 100 epochs, which means that the time consumed for an end-to-end training entirely in the encrypted domain would be prohibitive. Our idea is to take advantage of the fact that the feature vectors are in the clear, and hence it may be possible to decrypt the labels in each batch so that backpropagation can be carried out in cleartext, giving us computational advantage over an all encrypted solution. This is where we employ label differential privacy. A straightforward way to accomplish this is to let $P_2$ add differentially private noise to all its labels and simply handover its noisy dataset to $P_1$, playing no further part (except for receiving the accuracy result). However, this is less desirable from the utility point of view as we argue in  detail in Section~\ref{subsec:cipher-vs-plain}. Instead, we add noise to the gradients in each batch with $P_2$ interactively adding noise to the average gradient computed in each epoch, similar to what is done in~\cite{yuan2021label}.

\subsection{Proposed Protocol}
\label{subsec:protocol}
Our solution is as follows. Figure~\ref{fig:overview} shows the higher level overview of our protocol.
\begin{enumerate}
    \item To start, $P_2$ sends only the encrypted form of its dataset $D_2$ where each sample is $(\mathbf{x}, \llbracket \mathbf{y} \rrbracket_{\mathsf{k}})$, where $\mathsf{k}$ is $P_2$'s encryption key of a homomorphic encryption scheme. In particular, $\llbracket \mathbf{y} \rrbracket_{\mathsf{k}}$ is a vector of $K$ elements, each element of which is encrypted under $\mathsf{k}$.
    \item For each sample $s \in B$ and $1 \leq i \leq K$,  $P_1$ computes $\frac{\partial z_i(s)}{\partial \mathbf{w}}$. This results in $K \times |B|$ vectors of $R$-elements each, where $R$ is the number of weights in the last layer. 
    \item For each sample $s \in B$ and $1 \leq i \leq K$, $P_1$ does element-wise homomorphic scalar multiplication: $\frac{\partial z_i(s)}{\partial \mathbf{w}} \llbracket y_i(s) \rrbracket_{\mathsf{k}} = \llbracket y_i(s) \lfloor r \frac{\partial z_i(s)}{\partial \mathbf{w}} \rfloor \rrbracket_{\mathsf{k}}$, where $r$ is the precision parameter. This amounts to a total of $K \times |B| \times R$ homomorphic scalar multiplications. 
    \item $P_1$ homomorphically adds:
\begin{align}
\label{eq:ytimesz}
    & \sum_{s \in B} \sum_{i = 1}^K \left\llbracket y_i(s) \left\lfloor r \frac{\partial z_i(s)}{\partial \mathbf{w}} \right\rfloor \right\rrbracket_{\mathsf{k}} \nonumber\\
    = & \left\llbracket \sum_{s \in B} \sum_{i = 1}^K  y_i(s) \left\lfloor r \frac{\partial z_i(s)}{\partial \mathbf{w}} \right\rfloor \right\rrbracket_{\mathsf{k}}  = \left\llbracket N_B(\mathbf{w}) \right\rrbracket_\mathsf{k},
\end{align}
which amounts to a total of $K \times |B| \times R$ homomorphic additions. This results in an $R$-element vector encrypted under $\mathsf{k}$. 
\item $P_1$ computes the sensitivity of the gradients of the loss function for the current batch. As shown in Appendix~\ref{app:sense} this is $r \Delta S_B(\mathbf{w})$, where: 
\begin{equation}
    \label{eq:protocol-sensitivity}
    \Delta S_B(\mathbf{w}) = \frac{2}{|B|} \max_{i, s} \left\lVert   \frac{\partial z_i(s)}{\partial \mathbf{w}} \right\rVert_2,
\end{equation}
is the sensitivity of the gradients without encoding. 
\item $P_2$ computes $R$-dimensional Gaussian noise vector $\mathcal{N}(0, \sigma^2 \mathbf{1}_R)$, where 
$\sigma = 1/\epsilon$ and $\mathbf{1}_R$ is a vector of $R$ $1$'s. For each value $s$ in the list of allowable values of sensitivity of size $t$ (see below), $P_2$ updates the noise as $rs \mathcal{N}(0, \sigma^2 \mathbf{1}_R) = \mathcal{N}(0, (rs\sigma)^2 \mathbf{1}_R)$, and sends $\left\llbracket \lfloor \mathcal{N}(0, (rs\sigma)^2 \mathbf{1}_R) \rfloor \right\rrbracket_\mathsf{k}$ to $P_1$.
\item $P_1$ looks up the index of the smallest allowable sensitivity $s$ such that $\Delta S_B(\mathbf{w}) \leq s$. $P_1$ then chooses the noise vector  $\left\llbracket \lfloor \mathcal{N}(0, (rs\sigma)^2 \mathbf{1}_R) \rfloor \right\rrbracket_\mathsf{k}$ corresponding to this index as sent by $P_2$. 

\item $P_1$ ``blinds'' the encrypted quantity $\left\llbracket N_B(\mathbf{w}) \right\rrbracket_\mathsf{k}$ resulting in  $\left\llbracket N_B(\mathbf{w}) + \boldsymbol{\mu} \right\rrbracket_\mathsf{k}$ (see below). $P_1$ homomorphically adds the scaled noise vector and sends the following to $P_2$:
\begin{equation}
\label{eq:blinded}
    \left\llbracket N_B(\mathbf{w}) + \boldsymbol{\mu} + \lfloor \mathcal{N}(0, (rs\sigma)^2 \mathbf{1}_R) \rfloor \right\rrbracket_\mathsf{k}
\end{equation}
\item $P_2$ decrypts the ciphertext and sends the following to $P_1$
\begin{equation}
\label{eq:blinded-dec}
N_B(\mathbf{w}) + \boldsymbol{\mu} + \lfloor \mathcal{N}(0, (rs\sigma)^2 \mathbf{1}_R) \rfloor     
\end{equation}
\item $P_1$ substracts $\mu$ and obtains $N_B(\mathbf{w}) + \lfloor \mathcal{N}(0, (rs\sigma)^2 \mathbf{1}_R) \rfloor$. $P_1$ decodes this by dividing by $r$, plugs this into Equation~\eqref{eq:two-terms} and proceeds with backgpropagation in the unencrypted domain. 
\end{enumerate}

\begin{remark}
The parameter $R$, i.e., the number of weights in the last layer is being leaked here. We assume that this quantity, along with the batch size  and the number of epochs are known by party $P_2$.
\end{remark}

\begin{figure}
\centering
\begin{tikzpicture}

\node[] (phantom1) {};
\node[right=6.25cm of phantom1] (phantom2) {};

\node[minimum width = 3cm, minimum height = 7.8cm, below=-0.5cm of phantom1, fill=lgtgray, very thick, rounded corners] 		(entangle) 				{}; 

\node[minimum width = 1cm, minimum height = 7.8cm, below=-0.5cm of phantom2, fill=lgtgray, very thick, rounded corners] 		(entangle) 				{}; 

\node[] (P1) {$P_1$};
\node[right=6cm of P1] (P2) {$P_2$};

\node[draw=blue, fill=blue!2, right=2cm of P1] {Initialization};

\node[below=0.5cm of P1] (P1Init) {};
\node[below=0.5cm of P2] (P2Init) {};

\draw[<-] (P1Init) -- node [above] {Encrypted labels (\textsf{S1})} (P2Init);

\draw[dashed] ([xshift=-1cm, yshift=-0.2cm]P1Init.west) -- ([yshift=-0.2cm]P2Init.east);

\node[below=0.5cm of P1Init] (P1FP) {};
\node[below=0.5cm of P2Init] (P2FP) {};

\node[draw=blue, fill=blue!2, right=2.1cm of P1FP] {Forward Pass};

\node[below=0cm of P1FP, text width=3cm, align=center] () {Compute encrypted gradients (\textsf{S2-5})};

\draw[<-] ([yshift=-1.5cm]P1FP.south) -- node [above] {Encrypted noise list (\textsf{S6})} ([yshift=-1.5cm]P2FP.south);

\node[below=2cm of P1FP] () {Choose noise (\textsf{S7})};

\draw[->] ([yshift=-3cm]P1FP.south) -- node [above] {Encrypted noisy blinded gradients (\textsf{S8})} ([yshift=-3cm]P2FP.south);

\draw[<-] ([yshift=-3.5cm]P1FP.south) -- node [above] {Decrypted noisy blinded gradients (\textsf{S9})} ([yshift=-3.5cm]P2FP.south);

\node[below=4cm of P1FP,text width=3cm, align = center] () {Unblind and decode (\textsf{S10})};

\draw[dashed] ([xshift=-1cm, yshift=-5cm]P1FP.west) -- ([yshift=-5cm]P2FP.east);

\node[below=5.3cm of P1FP] (P1BP) {};
\node[below=5.3cm of P2FP] (P2BP) {};

\node[draw=blue, fill=blue!2, right=1.5cm of P1BP] {Backward Pass (in the clear)};

\end{tikzpicture}
\caption{\revise{Higher level overview of our protocol for one epoch for training the neural network. The quantity \textsf{S\#} in brackets indicates the step number in the protocol.}}
\label{fig:overview}
\end{figure}

\descr{Adding a Random Blind.} The labels are encrypted under party $P_2$'s key. Therefore, the computation of the quantity 
\[
\left\llbracket N_B(\mathbf{w}) + \lfloor \mathcal{N}(0, (rs\sigma)^2 \mathbf{1}_R) \rfloor \right\rrbracket_\mathsf{k}
\]
(without the blind) can be done homomorphically in the encrypted domain. At some point, this needs to be decrypted. However, decrypting this quantity will leak information about the model parameters to $P_2$. That's why $P_1$ uses the following construct to ``blind'' the plaintext model parameters. Namely, assume that encryption of a message $x$ is done under TLWE. Then before decryption, the ciphertext will be of the form:
\[
b = \langle \mathbf{s}, \mathbf{a} \rangle + x + e.
\]
$P_1$ samples an element $\mu$ uniformly at random from $\mathcal{P}$ and adds it to $b$. This then serves as a one-time pad, as the original message $m$ can be any of the $p$ possible messages in $\mathcal{P}$. Once the ciphertext has been decrypted, player $P_1$ receives $x + \mu$, from which $\mu$ can be subtracted to obtain $m$.

\descr{Allowable Values of Sensitivity.} The multiplication in Step 6 works for Gaussian noise because we can multiply a constant times a Gaussian distribution and still obtain a Gaussian with the scaled variance. If we encrypt the unit variance Gaussian noise first, then multiplication by a constant no longer implies a scaled Gaussian random variable, as the noise needs to be integer encoded before encryption. See Appendix~\ref{app:dp-violate}. As a result, we only use a predefined list of $t$ values of sensitivity. This incurs a slight utility cost, as $P_1$ would choose a value of sensitivity which is the smallest value greater than or equal to the true sensitivity in Eq.~\ref{eq:protocol-sensitivity}. Since $P_2$ does not know which of the $t$ values are used by $P_1$, we maintain secrecy of the actual sensitivity, albeit $P_2$ now knows that the sensitivity in each batch can only be one of the $t$ allowable values of sensitivity. For the concrete values used in our protocol, see Section~\ref{subsec:cipher-vs-plain}.

\subsection{Alternative Solutions}
Several other ways of constructing a protocol for the problem addressed in this paper are conceivable. However we discarded them as they had major shortcomings compared to our proposal. One solution is to have $P_2$ add differentially private noise to the labels of $D_2$, for instance via randomized response, and send them to $P_1$ playing no further part in the protocol. While computationally preferable, we address its limitations in terms of accuracy in detail in Section~\ref{subsec:cipher-vs-plain}. Another solution is for $P_2$ to train its own model on its dataset $D_2$, and then let $P_1$ query the trained model to check its consistency with $D_{\text{hold}}$. The major issue with this approach is that our scenario is expecting improvement over the combined dataset $D_1 \cup D_2$, and individually the datasets $D_1$ and $D_2$ may drift from $D_{\text{hold}}$. Furthermore, it relies on $P_2$'s expertise in model training. Another possibility is to simply assess the quality of the combined dataset through statistical tests. It is not clear what kind of and how many number of statistical tests would suffice to demonstrate that the machine learning model trained over the combined dataset would show improvement, let alone whether these tests can be efficiently performed in the encrypted domain. Thus, while other alternatives may exist, we believe that they are unlikely to provide substantial improvement over our proposal. 

\section{Privacy and Security Analysis}

\subsection{Proving Privacy}

Fix a batch $B$. The quantity $N_B(\mathbf{w})$ from Eq.~\ref{eq:ytimesz} is a vector of $R$ elements, $R$ being the number of weights in the last layer. This quantity is first encoded into the integer domain (for encryption). In Appendix~\ref{app:sense}, we show that the sensitivity of the loss function when no encoding is employed is given by:
\[
\lVert \nabla L_{B'}(\mathbf{w}) - \nabla L_{B''}(\mathbf{w}) \rVert_2  \leq \frac{2}{|B|} \max_{i, s} \left\lVert   \frac{\partial z_i(s)}{\partial \mathbf{w}} \right\rVert_2 = \Delta S_B(\mathbf{w}),
\]
where $B'$ and $B''$ are batches in neighbouring datasets $D'$ and $D''$. We then show that the sensitivity of the loss function when the partial derivative vector is encoded as $\left\lVert  \left\lfloor r \frac{\partial z_i(s)}{\partial \mathbf{w}} \right\rfloor \right\rVert_2$, as in the quantity $N_B(\mathbf{w})$, is given as: $r \Delta S_B(\mathbf{w})$.  
Let $rs \geq r \Delta S_B(\mathbf{w})$, where $s$ is the smallest value in the allowable list of sensitivity values that is greater than or equal to $\Delta S_B(\mathbf{w})$. Then the mechanism 
\[
 N_B(\mathbf{w})  + rs \mathcal{N}(0, \sigma^2 \mathbf{1}_R) =  N_B(\mathbf{w})  + \mathcal{N}(0, (rs\sigma)^2\mathbf{1}_R)
\]
is $\epsilon$-GLDP, where $\sigma = 1/\epsilon$, and $\mathbf{1}_R$ is an $R$-element vector each element of which is 1. To cast this result as an integer, we simply floor the result:
\[
\lfloor  N_B(\mathbf{w})  + \mathcal{N}(0, (rs\sigma)^2\mathbf{1}_R) \rfloor =  N_B(\mathbf{w}) + \lfloor \mathcal{N}(0, (rs\sigma)^2\mathbf{1}_R) \rfloor,
\]
as the noise is already of scale $\approx r$. The mechanism on the left remains differentially private due to the post-processing property of Gaussian differential privacy. The expression on the right remains differentially private due to the fact that $ N_B(\mathbf{w}) $ is an integer. In other words, we can generate noise and truncate it before adding it to the true value without losing the differential privacy guarantee, as is done in Steps 7 and 8 of the protocol. Since the batches are disjoint in each epoch (if not using a random sample), we retain $\epsilon$-GLDP by invoking parallel composition (Proposition~\ref{prop:seq-par-comp}). Over $n$ epochs the mechanism is $\sqrt{n} \epsilon$-GLDP invoking sequential composition (Proposition~\ref{prop:seq-par-comp}).   

\subsection{Proving Security}
Due to lack of space, we present the security proof in Appendix~\ref{app:security-proof} in the real-ideal paradigm~\cite[\S 23.5]{boneh2020graduate},\cite{canetti2001universal-composable}. Intuitively, $P_1$ only learns the noisy gradients of the batches, whereas $P_2$ does not learn parameters of $P_1$'s model except for batch size $|B|$, number of weights $R$ in the last layer, the number of epochs $n$, and the fact that the sensitivity can only be approximated to one of the $t$ allowable values of sensitivity.

\section{Experimental Evaluation}
In this section, we evaluate our protocol over several datasets. We first show that joining two datasets (in cleartext) does indeed improve the accuracy of the model. We then implement our protocol on Zama's \texttt{concrete} TFHE library~\cite{concrete}, show the training time and accuracy of the trained model, and the values of the privacy parameter $\epsilon$ where the model shows accuracy improvement over $P_1$'s dataset but less than the accuracy if the model were to train without differential privacy. This setting is ideal for $P_2$ as this would persuade $P_1$ to go ahead with the collaboration without revealing the fully improved model. Our code is available at \url{https://github.com/Ryndalf/Label-Encrypted}.

\descr{Configurations and Datasets.} For all the experiments, we used a 64-bit Ubuntu 22.04.2 LTS with 32G RAM and the 12th Gen Intel(R) Core(TM) i7-12700 CPU. We did not use GPUs for our experiments \revise{following existing works~\cite{hesamifard2018privacy,nandakumar2019towards,xu2019cryptonn,lou2020glyph,xu2021nn} to keep the comparison fair}. Table~\ref{tab:hyperparameters} shows the common hyperparameters used in our experiments. When using different values (e.g., number of neurons in the hidden layer), we attach them with the experimental results. Table~\ref{tab:datasets} illustrates the brief statistics (number of records, number of features, number of classes and number of samples in each class) of the datasets. All datasets contain (almost) balanced classes, except for Drebin \revise{and Purchase-10}, where \revise{some labels have about twice the number of samples than others.}

\begin{table}[!ht]
\centering
\caption{Common hyperparameters for all models.}
\label{tab:hyperparameters}
\begin{tabular}{l l}
\toprule
\textbf{Hyperparameters}& \textbf{Configuration}  \\
\midrule 
    activation function & sigmoid \\
    output function & softmax \\
    loss function & cross entropy \\ 
    optimiser & SGD \\ 
    $L2$ regulariser & 0.01 \\ 
\bottomrule
\end{tabular}
\end{table}

\begin{table}[!ht]
\centering
\caption{Datasets statistics.}
\label{tab:datasets}
\resizebox{\columnwidth}{!}{
\begin{tabular}{l|c|c|c|c}
\toprule
\textbf{Dataset} & \textbf{\#Rec.} & \textbf{\#Feat.} & \textbf{\#Classes} & \textbf{Class distribution}\\
\midrule
    Iris~\cite{uci-datasets} & 150 & 4 & 3 & Uniform Distribution\\ 
    Seeds~\cite{uci-datasets} & 210 & 7 & 3 & Uniform Distribution\\
    Wine~\cite{uci-datasets} & 178 & 13 & 3 & (59, 71, 58)\\
    Abrupto~\cite{concept-drift} & 10,000 & 4 & 2 & Uniform Distribution\\
    Drebin~\cite{drebin} & 16,680 & 3,506 & 2 & (5560, 11120)\\
    \revise{CIFAR-10}~\cite{cifar-datasets} & 60,000 & 3072 & 10 & Uniform Distribution\\
    \revise{CIFAR-100}~\cite{cifar-datasets} & 60,000 & 3072 & 100 & Uniform Distribution\\
    \revise{Purchase-10}~\cite{purchase-dataset} & 200,000 & 594 & 10 & {\thead{(15107, 17341, 22684, 11722, 12727 \\ 22257, 29432, 19919, 33842,14969)}}\\
\bottomrule
\end{tabular}
}
\end{table}

\subsection{Model Improvement by Joining Datasets}
\label{subsec:model-improve}
In Section~\ref{sec:accuracy-gain}, we explored if $P_2$ does not have domain knowledge, whether training the model $M_2$ on the dataset $D_1 \cup D_2$ will increase the accuracy of $M_2$ on $D_\text{hold}$, with $D_\text{hold}$ being a balanced dataset. 
Here, we are interested in knowing the other side of the coin: if $P_2$'s dataset $D_{2}$ is indeed of better quality, does this result in an improved performance on $M_2$? To demonstrate this, we will use the scenario where $D_1$ is small and imbalanced, i.e., for one of the labels, it has under-representative samples. On the other hand, $D_2$ is larger and more balanced. Intuitively, joining the two should show substantial improvement in accuracy.

\revise{To illustrate the effect, we use Abrupto dataset as a case study. In particular,} we use the first $10,000$ samples from the Abrupto dataset~\cite{concept-drift}, i.e., the dataset \texttt{mixed\_1010\_abrupto}, which is a balance binary dataset. 
From this dataset, we set aside 200 samples labelled 0 and 200 samples labelled 1 for $D_\text{hold}$, 96 samples labelled 0, and 864 samples labelled 1 to dataset $D_1$, and the remaining samples to dataset $D_2$. Note that $D_2$ is considerably larger and more balanced than dataset $D_1$. Given these datasets, model $M_1$ is trained on $D_1$, and model $M_2$ on $D_1 \cup D_2$, and their accuracies evaluated over $D_\text{hold}$. 

The configuration is the same for the two models except for the learning rate and batch sizes, which are tailored to account for the relative size difference between datasets $D_1$ and $D_1 \cup D_2$. With these settings, we report the average test accuracy (over $D_{\text{hold}}$) of training each model 10 times. 
These are shown in Table~\ref{tab:model-improve}. There is very little difference in accuracy results in each of these runs. Model $M_1$ achieves an average accuracy of $0.703$, whereas model $M_2$ achieves an accuracy of $0.912$. We therefore conclude that augmenting a small and unbalanced dataset with a large and balanced dataset will increase the accuracy of the resulting model (as long as the dataset labels are of good quality as ascertained by the holdout dataset).



\begin{table}[!th]
\centering
\caption{Accuracy of Models $M_1$ (hidden neurons: 4, batch size: 128, learning rate: 0.2, epochs: 100) and $M_2$ (hidden neurons: 4, batch size: 512, learning rate: 0.1, epochs: 100) against The Holdout Dataset.}
\label{tab:model-improve}
\begin{tabular}{c c c}
\toprule
\textbf{Run}& \textbf{Model  $M_1$ } &\textbf{Model $M_2$}\\
\midrule
    1& 0.700 & 0.912\\
    2& 0.700 & 0.915 \\
    3& 0.712 & 0.912\\
    4& 0.700 & 0.908 \\ 
    5& 0.700 & 0.915\\ 
    6& 0.700 & 0.912 \\
    7& 0.700 & 0.915\\ 
    8& 0.700 & 0.908\\ 
    9& 0.718 & 0.912\\
    10& 0.708 & 0.910 \\    
\midrule
   \textbf{Average}&0.704 & 0.912 \\
\bottomrule
\end{tabular}
\end{table}

\revise{However, as argued in Section~\ref{sec:accuracy-gain}, an imbalanced $D_1$ is risky in the sense that $D_1 \cup D_2$ might show improvement even when $P_2$ does not have any domain expertise. Thus, another, more appropriate scenario is when $D_1$ is smaller in size than $D_2$, but otherwise follows a similar label distribution as the true distribution. To see this, we also studied how the ratio of the number of samples of $D_{1}$ and of $D_{2}$ could impact the accuracy of the trained models $M_{1}$ and $M_{2}$. In general, a larger size difference between the two models, $D_1$ being smaller, shows an improvement in model accuracy. This is consistent with the results in Section~\ref{sec:accuracy-gain}. The experimental results are shown in Appendix~\ref{app:ratio}.}

\subsection{Neural Network Implementation}
There are plenty of libraries available for training deep neural networks, e.g., PyTorch. Looking at our protocol, we want to be able to access derivatives of the last layer in order to perform homomorphic encryption operations on them (Eq.~\eqref{eq:two-terms}). 
\revise{However, existing ML frameworks (e.g., Keras and PyTorch) return all calculated gradients after backpropagation completes. This means that for every epoch we need to wait until PyTorch completes backpropagation, pause it, retrieve all calculated gradients, modify the gradients in the last layer, and restart the completed backpropagation from the modified gradients. To save the run-time from these unnecessary steps,} 
we therefore implement the loss function, activation functions and their derivatives from scratch in Python 3.10 using \texttt{Numpy} and \texttt{sklearn} based on Eq.~\eqref{eq:two-terms}. To ensure our implementation (from scratch) ends up with the same learning outcome as the model implemented by \texttt{torch.nn} (the PyTorch neural network library), we compare the model parameters and the accuracy of the two models over the same training ($70\%$ of the raw Iris dataset~\cite{uci-datasets}) and test ($30\%$ of the raw Iris dataset~\cite{uci-datasets}) datasets. 

\descr{Training Outcomes.} For the two neural networks (excluding operations such as dropout), if the order in which the datasets are read, the initialization of the weights, and their parameter configurations (learning rate, epochs, etc.) are the same, then their final trained weights are also the same as can be seen from Table~\ref{tab:scratch-vs-actual} in Appendix~\ref{app:weights-biases}, where the weights assigned to each layer are exactly the same in the two implementations up to four decimal places of accuracy. \revise{Additionally, Figure~\ref{fig:ROC_from_scratch} in Appendix~\ref{app:weights-biases} further depicts the training performance measured by ROC of our model and the PyTorch implementation. As can be seen the two are almost identical. We are therefore convinced that our implementation from scratch is an accurate representation of the model from PyTorch.}


\descr{Implementation Time.} Table~\ref{tab:final_results3} compares the average (of 10 runs) training time (in seconds) of the model $M_2$ implemented from scratch (Model $M_2$, Plaintext, no DP), and the PyTorch baseline model (Model $M_{2}$ Baseline) on all datasets from Table~\ref{tab:datasets}. It is observed that, training time in the plaintext of our implementation is about 100 to 300 times slower than that of the PyTorch baseline. We analysed the source code of our implementation and the PyTorch documentation and then concluded three points caused this observation.
\begin{itemize}
    \item 
    Due to the need to incorporate Gaussian-distributed noise, smaller negative numbers may occur, which can lead to data overflow when the activation function sigmoid is entered. Therefore our program is designed in such a way that for all model training, the sigmoid will be adjusted to Eq.~\eqref{eq:sigmoid}.
    \begin{equation}
        \text{sigmoid}(x)=\begin{cases}
            \frac{1}{1+e^{-x}}, & \text{if}\ x \geq 0,\\
            \frac{e^{x}}{1+e^x}, & \text{otherwise.}\\
        \end{cases}
        \label{eq:sigmoid}
    \end{equation}
    
    \item 
    PyTorch optimizes low-level implementation for training, i.e., dynamic computational graphs, which means the network behaviour can be changed programmatically at training time to accelerate the training process.

    \item
    We implemented more functions to accommodate the \concrete{} library~\cite{concrete} even in the plaintext domain. A main deviation from the general neural network training is the way we treat the gradients. As discussed in Section~\ref{sec:intuition}, we calculate the gradients separately using two pathways depending on unencrypted (from $D_1$) vs encrypted labels (from $D_2$) as shown in Eq.~\eqref{eq:two-terms}. Hence, in our code, we defined a separate function for preparing a batch of sample forward propagation results, which are forced to be divided into two terms for computation.
\end{itemize}

\descr{Encrypted Domain.} Zama's current \texttt{Concrete} framework for TFHE~\cite{concrete} mandates decimals to be converted to integers. For our experiments, we chose a precision level of six decimal places, i.e., we use $r = 10^6$. Additionally, in order to verify that the encryption operations do not affect the accuracy of our implementation, we use the Iris dataset for a simple experiment. After setting the same initialization of weights and the order of reading the datasets, we train the model from scratch in plaintext and train another model from scratch on the ciphertext, and compare the trained model parameters. The weights of the two models were almost identical up to four decimal places. The exact weights and biases are shown in Table~\ref{tab:scratch-vs-encrypted} in Appendix~\ref{app:weights-plain-cipher}.

\begin{figure*}[!ht]
\centering
\resizebox{0.75\textwidth}{!}{
\begin{tikzpicture}

\node[minimum width = 8cm, minimum height = 5cm, draw=blue, fill=blue!2, rounded corners, very thick] 		(zama) 				{}; 

\node[] 		(zamalabel) 				[below left =0.5cm of zama.north east]{\Large \concrete{}}; 

\node[draw=black, minimum width = 1.5cm, minimum height = 0.5cm] 							(zw) 			[yshift=1.5cm, left=1cm of zama.west] {$z_\mathbf{w}$};

\node[draw=black, minimum width = 1.5cm, minimum height = 0.5cm] 							(zwr) 			[yshift=1.5cm, right=1cm of zama.west] {$z_\mathbf{w}$};

\draw[->] (zw.east) -- node [draw = black, fill = white, above=0.1em, pos = 0.5] {\footnotesize clear} (zwr.west);

\node[draw=black, minimum width = 1.5cm, minimum height = 0.5cm] 							(y) 			[yshift=0.5cm, left=1cm of zama.west] {$y$};

\node[draw=black, minimum width = 1.5cm, minimum height = 0.5cm, fill=gray, text=white] 							(yr) 			[yshift=0.5cm, right=1cm of zama.west] {$y$};

\draw[->] (y.east) -- node [draw = black, fill = white, above=0.1em, pos = 0.5] {\footnotesize encrypted} (yr.west);

\node[draw=black, minimum width = 1.5cm, minimum height = 0.5cm] 							(s) 			[yshift=-0.5cm, left=1cm of zama.west] {$rs \cdot \eta$};

\node[draw=black, minimum width = 1.5cm, minimum height = 0.5cm, fill=gray, text=white] 							(sr) 			[yshift=-0.5cm, right=1cm of zama.west] {$rs \cdot \eta$};

\draw[->] (s.east) -- node [draw = black, fill = white, above=0.1em, pos = 0.5] {\footnotesize encrypted} (sr.west);

\node[draw=black, minimum width = 1.5cm, minimum height = 0.5cm] 							(u) 			[yshift=-1.5cm, left=1cm of zama.west] {$\mu$};

\node[draw=black, minimum width = 1.5cm, minimum height = 0.5cm] 							(ur) 			[yshift=-1.5cm, right=1cm of zama.west] {$\mu$};

\draw[->] (u.east) -- node [draw = black, fill = white, above=0.1em, pos = 0.5] {\footnotesize clear} (ur.west);

\node[draw=black, circle, inner sep=0pt, fill=gray, text=white] 							(mul1) 			[right=2cm of yr.east, yshift=0.5cm] {$\times$};

\draw[->] (zwr.east) -- (mul1.north west);
\draw[->] (yr.east) -- (mul1.south west);

\node[draw=black, circle, inner sep=0pt, fill=gray, text=white] 							(plus) 			[right=6cm of zama.west, yshift=0cm] {$+$};

\draw[->] (mul1.south east) -- (plus.north west);
\draw[->] (sr.east) -- (plus.west);
\draw[->] (ur.east) -- (plus.south west);

\node[draw=black, align = center, text width=2.5cm] 							(noisy_g) 			[yshift=0cm, right=1cm of zama.east] {$N(\mathbf{w}) + rs \cdot \eta + \mu$};

\draw[->] (plus.east) -- node [draw = black, fill = white, above=0.1em, pos = 0.5] {\footnotesize decrypted} (noisy_g.west);

\node[draw=black, minimum width = 1.5cm, minimum height = 0.5cm] 							(u2) 			[yshift=0cm, right=1cm of noisy_g.east] {$\mu$};

\node[draw=black, circle, inner sep=0pt, align=center] 							(minus2) 			[below=0.5cm of noisy_g.south east, xshift=0.5cm] {$-$};

\node[draw=black, align = center, text width=2cm] 							(g) 			[below=0.5cm of minus2.south] {$N(\mathbf{w}) + rs \cdot \eta$};

\node[] () [below = 0.2cm of g.south] {\footnotesize DP gradients};

\draw[->] (noisy_g.south) -- (minus2.north west);
\draw[->] (u2.south) -- (minus2.north east);
\draw[->] (minus2.south) -- (g.north);

\end{tikzpicture}
}
    \caption{Our homomorphic encryption circuit in Zama's \concrete{}.}
    \label{fig:zama}
\end{figure*}

\subsection{Protocol Implementation}
We use the \concrete{} framework from Zama~\cite{concrete} to implement the TFHE components of our protocol. Note that not all the operations in our protocol require homomorphic operations. The circuit to compute the protocol operations for one sample is given in Figure~\ref{fig:zama}. Given a sample $s$ of the batch $B$, from Eq.~\eqref{eq:ytimesz} in Step 4 of the protocol, we first need to multiply the derivatives of the inputs to the softmax function with the encrypted labels. Dropping subscripts and abbreviating notation, these are shown as $z_\mathbf{w}$ and $y$ respectively in the figure. For brevity, we skip the integer encoding step, but it is understood that $z_\mathbf{w}$ is multiplied by $r$ (the precision parameter), truncated, and then multiplied with $y$ as given by Eq.~\eqref{eq:ytimesz}.  Next, Step 6 of the protocol samples noise of scale $\mathcal{N}(0, \sigma^2 \mathbf{1}_R)$, which for a single entry with Gaussian differential privacy is $\eta = \mathcal{N}(0, 1/\epsilon^2)$. 
The noise is then multiplied by the $t$ allowable values of sensitivity and then the truncated noise value is encrypted as in Step 6 of the protocol. Again to avoid excessive notation, we simply show this as multiplying $\eta$ by one of the allowable values of sensitivity $rs$ (noting $r$ being the precision parameter). These quantiities are then added homomorphically according to Step 8 of the protocol. On the other side of the circuit, once decrypted, we get the blinded and differentially private gradients (Eq~\eqref{eq:blinded-dec}), which result in the differentially private gradients after subtracting the blind.

\descr{Implementation Notes.} As \texttt{Concrete} only accepts integer inputs, we need to convert the respective inputs to integer equivalents. To do so, we compute $z_\mathbf{w}$ 
to six decimal places, i.e., we multiply it by the precision factor $r = 10^6$, and floor the result. Similarly, we floor $rs \cdot \eta$, since the noise is already of scale $r$. Thus, the two quantities are in effect multiples of $r = 10^6$. Note that the label $y$ is unchanged, as it is already an integer.

\subsection{Plaintext vs Ciphertext Versions of the Protocol}
\label{subsec:cipher-vs-plain}
To ensure that the protocol replicates the scenario of Section~\ref{subsec:model-improve}, we evaluate our protocol against the unencrypted setting. Namely, we evaluate model $M_1$ on dataset $D_1$, model $M_2$ on dataset $D_1 \cup D_2$, both unencrypted and without differential privacy, and model $\widetilde{M}_2$ on dataset $D_1 \cup D_2$ computed through our protocol (with encryption and differential privacy). We use all the datasets from Table~\ref{tab:datasets} to evaluate this. Furthermore, we retain $3,506$ features from the Drebin dataset which is a subset of all available features. The feature classes retained include `api\_call', `call', `feature', `intent', `permission', `provider', and `real\_permission'. 

For each dataset and each value of $\epsilon$, we report the average (of 10 runs) results. Each time, the dataset is re-partitioned and $30\%$ of the data is randomly used as $D_\text{hold}$. 
Due to the different sizes of the datasets, we divide $D_1$ and $D_2$ differently for different datasets:
\begin{itemize}
    \item For Iris, Seeds and Wine, $D_1$ is $10\%$ of the total data and $D_2$ is $60\%$ of the total data. 
    \item For Abrupto, Drebin, \revise{CIFAR-10, CIFAR-100 and Purchase-10,} $D_1$ is $1\%$ of the total data and $D_2$ is $69\%$ of the total data. 
\end{itemize}
Note that we do not artificially induce a skewed distribution of samples in $D_1$ for these experiments, as this was done in Section~\ref{subsec:model-improve}. 

Table~\ref{tab:final_results3} shows the results of our experiments. The column $\epsilon$ contains the overall privacy budget. We use $t = 100$ allowable values of sensitivity. In all our experiments, we found $\Delta S_B(\mathbf{w})$, i.e., the sensitivity of the gradients without encoding from Eq.~\eqref{eq:protocol-sensitivity}, ranged from $0.019$ to $0.035$. We extended this range to $[0, 0.1]$ and picked 100 evenly spaced samples as the $t = 100$ allowed values of sensitivity. For datasets Iris, Seeds and Wine, training our protocol ($\widetilde{M}_2$) is about 10,000 times slower than training the same model in the clear ($M_2$). However, this time is mainly due to the encryption of the noise vector with all $t = 100$ allowable values of sensitivity. The time per epoch and total time consumed in encrypting this list is shown in the column labelled ``$t$-list time'' in the table. The total time is obtained by multiplying per epoch time by 50, the total number of epochs. Subtracting this time means that these three datasets can be trained between 100 to 250 seconds, which is very reasonable. Note that this list of noisy sensitivities does not depend on the datasets. Therefore $P_2$ can precompute these lists, significantly reducing interactive time. With this speed-up our implementation is only $\approx$100 times slower than training in the clear, i.e., $M_2$. 
The $\epsilon = 100$ regime gives little to no privacy. Looking at the numbers corresponding to those rows, we see that the accuracy of $\widetilde{M}_2$ is almost identical to $M_2$. This shows that our encrypted portion of the protocol does not incur much accuracy loss. For each dataset, we also show values of $\epsilon$ between $0.25$ and $0.5$, where the accuracy of $\widetilde{M}_2$ lies nicely in between $M_1$ and $M_2$. Thus, this value of $\epsilon$ can be used by $P_2$ such that $\widetilde{M}_2$ shows an improvement over $M_1$, yet at the same time, training over the noiseless data promises even more improvement.

\begin{table*}[ht]
\centering
\caption{Training time and accuracy of $M_1$ (dataset $D_1$), $M_2$ (dataset $D_1 \cup D_2$) in the clear, $\widetilde{M}_2$ (dataset $D_1 \cup D_2$) through our protocol, $M^{\text{Py}}_2$ (dataset $D_1 \cup D_2$) implemented in PyTorch in the clear, and ${M}^{\text{RR}}_2$ (dataset $D_1 \cup D_2$) with randomized response (hidden neurons: 20, batch size: 256, learning rate: 0.1, epochs: 50, weight decay: 0.01).}
\label{tab:final_results3}
\resizebox{\textwidth}{!}{
\begin{tabular}{c|c| c c| c c | c | c c |c c | c}
\toprule
\multirow{3}{*}{Dataset} & \multirow{2}{*}{$\epsilon$}  &  \multicolumn{2}{c|}{\thead{Model $M_1$ \\ Plaintext, no DP}} & \multicolumn{2}{c|}{\thead{Model $M_2$\\ Plaintext, no DP}} &  \multicolumn{3}{c|}{\thead{Model $\widetilde{M}_2$\\ Our protocol}} & \multicolumn{2}{c}{\thead{Model $M^{\text{Py}}_2$\\ PyTorch baseline}} &\multicolumn{1}{|c}{\thead{Model $M^{\text{RR}}_2$ \\ Randomized Response}}  \\
\cline{3-12}
 & & Time (s) & Test Acc. & Time (s) & Test Acc. & \thead{ $t$-list time (s) \\ per epoch \& total} &Gap Time (s) & Test Acc. & Time (s) & Test Acc.  & Test Acc.\\ 
\hline
\multirow{7}{*}{Iris} 
&0.1 &\multirow{7}{*}{0.0619}&\multirow{7}{*}{0.7289}&\multirow{7}{*}{0.4244}&\multirow{7}{*}{0.8467}& \multirow{6}{*}{\shortstack[c]{1m11s\\ $\approx$58m57s}} &  10m55s &  0.7733 &\multirow{7}{*}{0.0078}&\multirow{7}{*}{0.8467} &0.3711 \\ 
&1 &&&&&  &11m13s  &  0.8511  &&& 0.6244\\
&10 &&&&&   & 10m52s     &  0.8521  &&& 0.8189\\
&100 &&&&&   & 10m49s     &  0.8554 &&& 0.8311\\
\cline{2-2}\cline{8-9}
&0.2&&&&&    &  11m14s   &  0.7821  &&&0.3416\\
&0.5&&&&&    &  11m11s   &  0.8422  &&&0.6266\\\hline

\multirow{7}{*}{Seeds}
&0.1&\multirow{7}{*}{0.0896}&\multirow{7}{*}{0.8079}&\multirow{7}{*}{0.6216}&\multirow{7}{*}{0.8762}&  \multirow{6}{*}{\shortstack[c]{1m38s\\ $\approx$1h21m55s}} & 13m19s  &  0.6132  &\multirow{7}{*}{0.0084}&\multirow{7}{*}{0.8762}& 0.3682\\
&1&&&&&   & 13m7s     &  0.8762  &&& 0.6873\\
&10&&&&&  &  13m24s     &  0.8719  &&& 0.8550\\
&100&&&&&   & 12m54s     &  0.8730  &&& 0.8762\\
\cline{2-2}\cline{8-9}
&0.2&&&&&   &  12m35s    &  0.8111  &&&0.3794\\
&0.5&&&&&   &  13m36s    &  0.8714  && &0.6174\\\hline

\multirow{7}{*}{Wine} 
&0.1&\multirow{7}{*}{0.1161}&\multirow{7}{*}{0.7981}&\multirow{7}{*}{0.5265}&\multirow{7}{*}{0.9302}&   \multirow{6}{*}{\shortstack[c]{01m24s\\ $\approx$1h10m12s}} &  12m59s   &  0.7792  &\multirow{7}{*}{0.0105}&\multirow{7}{*}{0.9302} & 0.4361\\
&1 &&&&&  & 13m17s      &  0.9340  && & 0.7981\\
&10&&&& &  & 13m5s      &  0.9396  &&& 0.9205\\
&100&&&& & &  12m56s      &  0.9396  &&& 0.9283\\
\cline{2-2}\cline{8-9}
&0.2&&&& &   &  13m2s    &  0.8905  &&&0.4999\\
&0.5&&&& &   & 12m49s     &   0.9320 &&&0.5168\\\hline


\multirow{6}{*}{Abrupto}
&0.1&\multirow{6}{*}{0.4212}&\multirow{6}{*}{0.8399}&\multirow{6}{*}{28.3430}&\multirow{6}{*}{0.9077}&  \multirow{6}{*}{\shortstack[c]{53m15s\\ $\approx$1d20h22m38s}} & 4h27m20s & 0.7794   &\multirow{6}{*}{0.2687}&\multirow{6}{*}{0.9077}& 0.6593\\
&1 &&&& &  & 4h27m5s      &  0.9028  &&& 0.8869\\
&10 &&&& &  & 4h27m14s      &  0.9059  &&& 0.9004\\
&100 &&&& &  & 4h27m6s       &  0.9061  &&& 0.9067\\
\cline{2-2}\cline{8-9}
&0.2 &&&& & & 4h27m3s   & 0.8850  &&& 0.7413\\
&0.5&&&& & &4h27m13s     & 0.9045 &&&0.8916 \\\hline

\multirow{6}{*}{Drebin} 
&0.1 &\multirow{6}{*}{8.5758}&\multirow{6}{*}{0.8571}&\multirow{6}{*}{10m28s}&\multirow{6}{*}{0.9470}&  \multirow{6}{*}{\shortstack[c]{1h11m55s\\ $\approx$2d11h56m5s}}& 23h09m8s  &  0.7010  &\multirow{6}{*}{2.8677}&\multirow{6}{*}{0.9470}& 0.6053\\
&1 &&&& &  &23h09m4s       &  0.9235  &&&0.9191\\
&10 &&&& &  &23h09m20s       &  0.9462  &&& 0.9490\\
&100&&&& &  &23h09m12s       &  0.9476  &&& 0.9489\\
\cline{2-2}\cline{8-9}
&0.5&&&& &  &23h09m6s       &  0.8860 &&&0.8741\\ 
&0.7&&&& &  & 23h09m13s      & 0.9172  &&&0.9023\\
\bottomrule
\end{tabular}
}

\end{table*}

The appropriate choice of $\epsilon$ depends on the size of the dataset, as it is different for the two larger datasets. One is a synthetic dataset Abrupto and the other one is a real-world Android malware dataset Drebin. Table~\ref{tab:final_results3} also shows the results of these two larger datasets. In terms of training time, unlike small datasets, there is a significant difference in their training time. This is due to the large number of features in the Drebin dataset, which is around 876 times more than the features in the Abrupto dataset. However, once again, the majority of the time is consumed in encrypting the noisy $t$-list, which as we mentioned before, can be precomputed. Since these datasets have multiple batches per epoch, the per epoch time for the $t$-list is also higher. However, subtracting the total time of the $t$-lists, the training of these two datasets takes between 2,000 to 11,000 seconds, which is reasonably fast considering the sizes of these datasets. When $\epsilon=100$, the accuracy of $\widetilde{M}_2$ is almost the same or even better than that of $M_2$. For Abrupto datasets, at $\epsilon \in $ [0.15, 0.5] we find the spot between the accuracy of model $M_1$ and $M_2$. On the other hand, with $\epsilon \in [0.3, 0.5]$ we find the spot between the accuracy of model $M_1$ and $M_2$ based on the Drebin dataset.

\descr{Randomized Response-based Protocol.} In Section~\ref{sec:intuition} we briefly mentioned that one way to obtain an updated model $M_2$ is for $P_2$ to apply differentially private noise to the labels of $D_2$, and then handover the noisy labelled dataset to $P_1$. This will then enable $P_1$ to run the entire neural network training in the clear, giving significant boost in training time. This can be done by adding noise to the labels via the randomized response (RR) protocol, given in~\cite{balle2019shuffle}, which is used in many previous works as well. However, our experiments show that for a reasonable guarantee of privacy of the labels, the resulting accuracy through RR is less than the accuracy of even $M_1$ in most cases. To be more precise, let us detail the RR protocol for binary labels. For each label, $P_2$ samples a bit from a Bernoulli distribution with parameter $\gamma$. If the bit is 0, $P_2$ keeps the true label, otherwise, $P_2$ replaces the label with a random binary label~\cite{balle2019shuffle}. Note that the probability that the resulting label is the true label is $1 - \gamma/2$. The mechanism is $\epsilon$-label differentially private as long as:
\begin{equation}
\label{eq:rr}
    \frac{1 - \gamma/2}{\gamma/ 2} \leq {e^{\epsilon}} \Rightarrow \epsilon \geq \ln (\gamma/2 - 1)
\end{equation}
for all $\gamma < 1$. Setting the value of $\epsilon$ exactly equal to the quantity on the right, we see that the probability that the resulting label is the true label is given by $e^{\epsilon}/(e^{\epsilon} + 1)$. Thus, for instance, if $\epsilon = 1$, the probability of the label being the true label is $\approx 73\%$, for $\epsilon = \ln 3$ it is $75\%$, for $\epsilon = 3$ it is $\approx 95\%$, for $\epsilon = 5$ it is $\approx 99.33\%$, and for $\epsilon = 10$ it is $\approx 99.99\%$. Thus, for reasonable privacy we would like to choose $\epsilon$ less than 1, and definitely not close to 10, otherwise, $P_2$ is effectively handing over the true labels to $P_1$.

We trained the model by using the above mentioned RR protocol. The results are shown in Table~\ref{tab:final_results3} under the column labelled ${M}^{\text{RR}}_2$. As we can see, the accuracy is less than the accuracy of even $M_1$ for the smaller Iris, Seeds and Wine datasets, only approaching the accuracy between $M_1$ and $M_2$ when $\epsilon$ is close to $10$, but which as we have argued above is not private enough. For the two larger datasets, Abrupto and Drebin, $\epsilon = 1$ gives better accuracy than $M_1$ while being less than $M_2$. However, we achieve the same using our protocol for a significantly smaller value of $\epsilon$. This is advantageous because it means that $P_1$ and $P_2$ can do multiple collaborations without exhausting the budget. 

\descr{Effect of the Number of Classes and Dataset Size on Runtime.} \revise{The analogous results on the three larger datasets, CIFAR-10, CIFAR-100 and Purchase-10 to the ones shown in Table~\ref{tab:final_results3} are shown in the Appendix~\ref{app:runtime:cifar} due to lack of space, where we evaluate the effect of the size of the dataset and the number of classes on the runtime.}

\descr{Communication Overhead.} The TLWE ciphertext is defined as $\mathbf{c} = (\mathbf{a}, b) \in \mathbb{T}_{q}^{N+1}$. As mentioned in Section~\ref{subsec:bg}, we use $N = 630$ and $q$ is 64 bits long. This means that the ciphertext is of size: $631 \times 64 = 40,384$ bits. Luckily, as shown in~\cite{joye2021guide}, a more compact way to represent $\mathbf{c}$ is to first uniformly sample a random seed $\theta$ of 128 bits (the same level of security as the key $\mathsf{k}$), and then use a cryptographically secure PRNG to evaluate the vectors $\mathbf{a} \leftarrow \text{PRNG}(\theta)$. This ciphertext representation is only $128 + 64 = 192$ bits long (adding the bit-lengths of $\theta$ and $b$). This reduction applies only when $P_2$ generates fresh ciphertexts, as any homomorphic computations will, in general, not be able to keep track of the changes to the vectors $\mathbf{a}$. With these calculations, we estimate the communication cost of our protocol as follows:
\begin{itemize}
    \item In Step 1 of the protocol, $P_2$ sends $m_2$ encrypted labels under key $\mathsf{k}$ of its dataset $D_2$ of size $m_2$. Using the compact representation just mentioned, this results in communication cost of $192m_2$ bits. We ignore the cost of plaintext features of $D_2$ being transmitted, as this cost is the same for any protocol, secure or not.
    \item In Step 6, $P_2$ sends the encryption of the $R$-element noise vector for all $t$ allowable values of sensitivity. Once again, since these are fresh encryptions under $\mathsf{k}$, we have communication cost of $192tR$ bits.
    \item In Step 8, $P_1$ sends the $R$-element encrypted, blinded and noise added vector to $P_2$. Since this is obtained through homomorphic operations, we use the non-compact representation, yielding $40384R$ bits of communication cost.
    \item Lastly, in Step 9 $P_2$ sends the decrypted version of the quantity above. The communication cost is less than $64R$ bits, as this is plaintext space. 
\end{itemize}
Note that Step 6, 8 and 9 need to be repeated for every batch in an epoch. Let $m_{1, 2}$ be the size of the dataset $D_1 \cup D_2$. Then the number of times these steps are repeated in each epoch is given by $\lceil m_{1, 2}/ |B| \rceil$, where $|B|$ is the batch size. Thus the total communication cost for one epoch is $192m_2 + (192tR + 40384R + 64R)(\lceil m_{1, 2}/ |B| \rceil = 192m_2 + (192tR + 40448R)(\lceil m_{1, 2}/ |B| \rceil$. Over $n$ epochs, the communication cost in bits is thus:
\begin{equation}
\label{eq:comm-cost}
C = \left(192m_2 + \left(192t + 40448\right)R\left\lceil \frac{m_{1,2}}{|B|}\right\rceil\right)n
\end{equation}
For the datasets in Table~\ref{tab:final_results3}, we have $R = 40$, since the number of hidden neurons is 20 (and binary classification), number of epochs $n = 50$, and $t = 100$. The size of $D_2$, i.e., $m_{2}$ is $0.6m$ for Iris, Seeds and Wine datasets, and equals $0.69m$ for the remaining two datasets where $m$ is the total size of the datasets as mentioned in Section~\ref{subsec:cipher-vs-plain}. For all datasets we have $m_{1, 2} = 0.7m$. The batch size $|B| = 256$.  The communication cost in units of megabytes MB is shown in Table~\ref{tab:comm-cost}. By far, the major cost is the $t$ encrypted noise values which need to be sent for each batch in an epoch in case of larger datasets. As mentioned earlier, these encrypted noise vectors can be generated well in advance to reduce interactive communication cost, as they do not use of any data specific parameter. For instance, with precomputed noise lists, the communication cost on the Drebin dataset reduces to 9.54MB per epoch. These times without the encrypted noise vectors are shown in the last two columns of the table labelled ``no $t$-list.'' Regardless, we see that communication in each round of the protocol (each epoch) can be done in a few seconds with typical broadband speeds.

\begin{table}[!ht]
\centering
\caption{Communication cost in megabytes (MBs) of our protocol as calculated through Eq.~\ref{eq:comm-cost} for the five datasets used in our implementation.}
\label{tab:comm-cost}
\resizebox{\columnwidth}{!}{
\begin{tabular}{l|c|c|c|c|c}
\toprule
\multirow{2}{*}{\textbf{Dataset}} & {\textbf{Size}} & \multirow{2}{*}{\textbf{Cost per epoch}} & \multirow{2}{*}{\textbf{Total cost}} & \textbf{Cost per epoch} & {\textbf{Total cost}}\\
& $m$ & & & no $t$-list & no $t$-list\\
\midrule
    Iris & 150 & 1.16MB & 58.22MB & 0.20MB & 10.22MB\\ 
    Seeds & 210 & 1.17MB & 58.26MB & 0.21MB & 10.26MB\\
    Wine & 178 & 1.16MB & 58.24MB & 0.20MB & 10.24MB\\
    Abrupto & 10,000 & 32.69MB & 1,634.34MB& 5.81MB & 290.33MB\\
    Drebin & 16,680 & 53.70MB & 2,685.16MB & 9.54MB & 477.16MB\\
\bottomrule
\end{tabular}
}
\end{table}

\subsection{Discussion -- Privacy Budget, Training Time and Limitations}
\label{sub:limitations}
\descr{Setting $\epsilon$ for Data Collaboration.} The value of $\epsilon$ used depends on the size of the dataset, with a smaller $\epsilon$ required for larger datasets. However, for all the datasets used in the experiment, with $\epsilon$ in the range $(0.3,0.5)$ the accuracy of the model $\widetilde{M_2}$ is higher than that of $M_1$, and lower than that of $M_2$. Thus, party $P_2$ can set an $\epsilon$ within this range, which provides visible benefits for $P_{1}$ to proceed with data collaboration in the clear. 

\descr{Fast Training of Our Protocol.} The training time of our protocol (Table~\ref{tab:final_results3}) is many orders of magnitude faster than protocols using entirely FHE operation reported in \cite{hesamifard2018privacy,nandakumar2019towards,lou2020glyph}. This is a direct result of keeping the forward and backward propagation in the clear even though some labels from the training data are encrypted. To compare the training time of our protocol against an end-to-end encrypted solution for neural network training, we look at the work from~\cite{hesamifard2018privacy}. They use polynomial approximations for the activation functions (e.g., sigmoid) to allow homomorphic operations. In one set of experiments, they train a neural network with one hidden layer over the Crab dataset, which has 200 rows and two classes. This dataset is comparable in size and number of classes to the Iris dataset used by us. They implement the homomorphically encrypted training of the neural network using HELib~\cite{halevi2020helib}. The time required for training one batch per epoch is 217 seconds (Table~3a in~\cite{hesamifard2018privacy}). For 50 epochs, and processing all batches in parallel, this amounts to 10,850 seconds. We note that after each round the ciphertext is sent to the client to re-encrypt and send fresh ciphertexts back to the server in order to reduce noise due to homomorphic operations. Thus, this time is a crude lower bound on the total time. In comparison, our protocol takes a total time of around 4,200 seconds for the entire 50 epochs over the comparable Iris dataset with one hidden layer, and just 150 seconds if the noise lists are pre-processed. (Table ~\ref{tab:final_results3}). With a more optimised implementation (say via PyTorch), this time can be reduced even further. 

\descr{Implementation Limitations and Potential Speedups.} The minimum value of the total budget $\epsilon$ tested by us is $0.1$. This is because of the limitation of \concrete{} in handling high precision real numbers (as they need to be converted into integers). When we inject differentially private noise using a (very) small $\epsilon$, it is highly likely to be inaccurate as it generate numbers with large absolute value, leading to the float overflow problem for sigmoid. In addition to that, our implementation of a neural network is many orders of magnitude slower than the PyTorch benchmark even though our accuracy matches that of the PyTorch baseline. If we are able to access gradients in a batch from the PyTorch implementation, then we would significantly accelerate the proposed protocol. \revise{For instance, the model ${M}_2$ through our implementation takes about 220 magnitudes more time than the baseline $M^{\text{Py}}_2$ through PyTorch on the Drebin dataset, i.e., $10$ mins $28$ seconds versus only 2.8677 seconds (see Table~\ref{tab:final_results3}). Thus our protocol on this dataset can potentially run in just over $6$ mins if run over PyTorch's implementation which is significantly lower than the 23 hours taken by current implementation. Furthermore, computing mini-batches in parallel can further improve computational time especially if run over GPUs. Lastly, as mentioned above the encrypted values of noise in the $t$-list can be pre-computed.}

\descr{Malicious Parties.} \revise{Our protocol is only secure in the honest-but-curious model. In the malicious setitng, many new challenges arise. A key challenge is to maintain efficiency as several constructs used by us cannot be used in the malicious setting, e.g., the use of the random blind. Another issue in the malicious model is that $P_1$ may lie about how many samples in the current batch are from $P_2$'s dataset and hence have less noise added to the gradients. We therefore leave protocol for the malicious setting as future work. 
}

\section{Related Work}
The closest work to ours is that of Yuan et al~\cite{yuan2021label}. They assume a scenario where one of the two parties holds the feature vectors and the labels are secret shared between the two parties. The goal is to jointly train a neural network on this dataset. At the end of the protocol, the first party learns the trained model whereas the second party does not learn anything. Like us, they use label differential privacy to obtain a more computationally efficient solution. Our scenario is different in that it enables party $P_1$ to assess the quality of the resulting dataset \emph{without knowing} the model, where $P_1$ does not initially trust the labelling from $P_2$. Another major difference is that we use FHE instead of secure multiparty computation (SMC) to train the model~\cite{yuan2021label}. Most importantly, it is unclear whether their protocol remains differentially private as they also encode the noise in the group of integers modulo a prime before multiplying with encoded sensitivity. As mentioned in Section~\ref{subsec:protocol}, this multiplication no longer means that the resulting Gaussian is appropriately scaled, resulting in violation of differential privacy. We show this in detail in Appendix~\ref{app:dp-violate}. It is unclear how their protocol can be modified to include pre-computed shares of a list of allowable sensitivity values as we do to resolve this issue. \revise{For instance, one party multiplies all sensitivities with unit Gaussian noise in the allowable list of sensitivites. In order to hide the actual noise, the resulting list needs to be secret shared with the second party. However, the second party is the one which knows the actual sensitivity. Hence telling the first party which share to use will reveal the sensitivity of the gradients.} 

Apart from Yuan et al~\cite{yuan2021label}, several works have investigated training neural network models entirely in the encrypted domain, encompassing both features and labels~\cite{hesamifard2018privacy,nandakumar2019towards,xu2019cryptonn,lou2020glyph,xu2021nn}. Hesamifard et al~\cite{hesamifard2018privacy} propose the CryptoDL framework, which employs Somewhat Homomorphic Encryption (SWHE) and Leveled Homomorphic Encryption (LHE) on approximated activation functions to facilitate interactive deep neural network training over encrypted training sets. However, the training time is prolonged even on small datasets (e.g., on the Crab dataset with dimensions of $200 \times 6$ cells, it takes over $200$ seconds per epoch/iteration during the training phase). Furthermore, the omission of details regarding the CPU clock speed (frequency) and the number of hidden layer neurons in their experiments renders the reported training time less equitable. Nandakumar et al~\cite{nandakumar2019towards} introduce the first fully homomorphic encryption (FHE)-based stochastic gradient descent technique (FHE-SGD). As pioneers in this field, FHE-SGD investigates the feasibility of training a DNN in the fixed-point domain. Nevertheless, it encounters substantial training time challenges due to the utilisation of BGV-lookup-table-based sigmoid activation functions. Lou et al~\cite{lou2020glyph} present the Glyph framework, which expedites training for deep neural networks by alternation between TFHE (Fast Fully Homomorphic Encryption over the Torus) and BGV~\cite{brakerski2014leveled} cryptosystems. However, Glyph heavily relies on transfer learning to curtail the required training epochs/iterations, significantly reducing the overall training time for neural networks. It should be noted that Glyph's applicability is limited in scenarios where a pre-trained teacher model is unavailable. Xu et al propose CryptoNN~\cite{xu2019cryptonn}, which employs functional encryption for inner-product~\cite{abdalla2015simple} to achieve secure matrix computation. However, the realisation of secure computation in CryptoNN necessitates the presence of a trusted authority for the generation and distribution of both public and private keys, a dependency that potentially compromises the security of the approach. NN-emd~\cite{xu2021nn} extends CryptoNN's capabilities to support training a secure DNN over vertically partitioned data distributed across multiple users.

By far, the major bottleneck of end-to-end homomorphic encryption or functional encryption approaches to neural network training is the runtime. For instance, a single mini-batch of 60 samples can take anywhere from 30 seconds to several days with dedicated memory~\cite{xu2021nn}. Another option is to use a secure multiparty computation (SMC) approach. In this case, the (joint) dataset can be secret shared  between two parties, and they can jointly train the model learning only the trained model~\cite{mohassel2017secureml}. However, once again end-to-end SMC solutions remain costly. For instance, the work from~~\cite{mohassel2017secureml} achieves SMC-based neural network training in 21,000 seconds with a network of 3 layers and 266 neurons. Undoubtedly, these benchmarks are being surpassed, e.g.,~\cite{keller2022secure}, however, it is unlikely that they will achieve the speed achieved by our solution, or the one from Yuan et al.~\cite{yuan2021label}. One way to improve our work would be to explore a combination of FHE and SMC. 


A related line of work looks at techniques to check the validity of inputs without revealing them. For instance, one can use zero-knowledge range proofs~\cite{morais2019zkrp} to check if an input is within an allowable range without revealing the input, e.g., age. This has, for instance, been used in privacy-preserving joint data analysis schemes such as Drynx~\cite{Froelicher2019} and Prio~\cite{corrigan2017prio} to ensure that attribute values of datasets are within the allowable range. However, in our case, we do not assume that the party $P_2$ submits any label that is out of range. Instead, the party may not have the domain expertise to label feature vectors correctly. This cannot be determined through input validity checking.

\section{Conclusion}
We have shown how two parties can assess the value of their potential machine learning collaboration without revealing their models and respective datasets. With the use of label differential privacy and fully homomorphic encryption over the torus, we are able to construct a protocol for this use case which is many orders of magnitude more efficient than an end-to-end homomorphic encryption solution. Our work can be improved in a number of ways. Due to several limitations in accessing components in PyTorch's neural network implementation and the integer input requirement in Zama's \concrete{} TFHE framework, our implementation is short of speedups that can potentially be achieved. As a result, we are also not able to test our protocol on larger datasets (say, 100k or more rows). Future versions of this framework may remove these drawbacks. Finally, there could be other ways in which two parties can check the quality of their datasets. We have opted for the improvement in the model as a proxy for determining the quality of the combined dataset. 

\bibliographystyle{ACM-Reference-Format}
\bibliography{enc_label.bib}


\begin{thebibliography}{36}


\ifx \showCODEN    \undefined \def \showCODEN     #1{\unskip}     \fi
\ifx \showDOI      \undefined \def \showDOI       #1{#1}\fi
\ifx \showISBNx    \undefined \def \showISBNx     #1{\unskip}     \fi
\ifx \showISBNxiii \undefined \def \showISBNxiii  #1{\unskip}     \fi
\ifx \showISSN     \undefined \def \showISSN      #1{\unskip}     \fi
\ifx \showLCCN     \undefined \def \showLCCN      #1{\unskip}     \fi
\ifx \shownote     \undefined \def \shownote      #1{#1}          \fi
\ifx \showarticletitle \undefined \def \showarticletitle #1{#1}   \fi
\ifx \showURL      \undefined \def \showURL       {\relax}        \fi
\providecommand\bibfield[2]{#2}
\providecommand\bibinfo[2]{#2}
\providecommand\natexlab[1]{#1}
\providecommand\showeprint[2][]{arXiv:#2}

\bibitem[Abdalla et~al\mbox{.}(2015)]%
        {abdalla2015simple}
\bibfield{author}{\bibinfo{person}{Michel Abdalla}, \bibinfo{person}{Florian
  Bourse}, \bibinfo{person}{Angelo De~Caro}, {and} \bibinfo{person}{David
  Pointcheval}.} \bibinfo{year}{2015}\natexlab{}.
\newblock \showarticletitle{Simple functional encryption schemes for inner
  products}. In \bibinfo{booktitle}{\emph{IACR International Workshop on Public
  Key Cryptography}}. \bibinfo{publisher}{Springer}, \bibinfo{pages}{733--751}.
\newblock


\bibitem[Arp et~al\mbox{.}(2014)]%
        {drebin}
\bibfield{author}{\bibinfo{person}{Daniel Arp}, \bibinfo{person}{Michael
  Spreitzenbarth}, \bibinfo{person}{Malte Hubner}, \bibinfo{person}{Hugo
  Gascon}, \bibinfo{person}{Konrad Rieck}, {and} \bibinfo{person}{CERT
  Siemens}.} \bibinfo{year}{2014}\natexlab{}.
\newblock \showarticletitle{Drebin: Effective and explainable detection of
  android malware in your pocket.}. In \bibinfo{booktitle}{\emph{Ndss}},
  Vol.~\bibinfo{volume}{14}. \bibinfo{pages}{23--26}.
\newblock


\bibitem[Balle et~al\mbox{.}(2019)]%
        {balle2019shuffle}
\bibfield{author}{\bibinfo{person}{Borja Balle}, \bibinfo{person}{James Bell},
  \bibinfo{person}{Adri{\`a} Gasc{\'o}n}, {and} \bibinfo{person}{Kobbi
  Nissim}.} \bibinfo{year}{2019}\natexlab{}.
\newblock \showarticletitle{The privacy blanket of the shuffle model}. In
  \bibinfo{booktitle}{\emph{Advances in Cryptology--CRYPTO 2019: 39th Annual
  International Cryptology Conference, Santa Barbara, CA, USA, August 18--22,
  2019, Proceedings, Part II 39}}. Springer, \bibinfo{pages}{638--667}.
\newblock


\bibitem[Blum and Hardt(2015)]%
        {blum2015ladder}
\bibfield{author}{\bibinfo{person}{Avrim Blum} {and} \bibinfo{person}{Moritz
  Hardt}.} \bibinfo{year}{2015}\natexlab{}.
\newblock \showarticletitle{The ladder: A reliable leaderboard for machine
  learning competitions}. In \bibinfo{booktitle}{\emph{International Conference
  on Machine Learning}}. PMLR, \bibinfo{pages}{1006--1014}.
\newblock


\bibitem[Boneh and Shoup(2023)]%
        {boneh2020graduate}
\bibfield{author}{\bibinfo{person}{Dan Boneh} {and} \bibinfo{person}{Victor
  Shoup}.} \bibinfo{year}{2023}\natexlab{}.
\newblock \showarticletitle{A graduate course in applied cryptography}.
\newblock \bibinfo{journal}{\emph{Draft 0.6}} (\bibinfo{year}{2023}).
\newblock


\bibitem[Brakerski et~al\mbox{.}(2014)]%
        {brakerski2014leveled}
\bibfield{author}{\bibinfo{person}{Zvika Brakerski}, \bibinfo{person}{Craig
  Gentry}, {and} \bibinfo{person}{Vinod Vaikuntanathan}.}
  \bibinfo{year}{2014}\natexlab{}.
\newblock \showarticletitle{(Leveled) fully homomorphic encryption without
  bootstrapping}.
\newblock \bibinfo{journal}{\emph{ACM Transactions on Computation Theory
  (TOCT)}} \bibinfo{volume}{6}, \bibinfo{number}{3} (\bibinfo{year}{2014}),
  \bibinfo{pages}{1--36}.
\newblock


\bibitem[Canetti(2001)]%
        {canetti2001universal-composable}
\bibfield{author}{\bibinfo{person}{Ran Canetti}.}
  \bibinfo{year}{2001}\natexlab{}.
\newblock \showarticletitle{Universally composable security: A new paradigm for
  cryptographic protocols}. In \bibinfo{booktitle}{\emph{Proceedings 42nd IEEE
  Symposium on Foundations of Computer Science}}. IEEE,
  \bibinfo{pages}{136--145}.
\newblock


\bibitem[Chaudhuri and Hsu(2011)]%
        {label-dp}
\bibfield{author}{\bibinfo{person}{Kamalika Chaudhuri} {and}
  \bibinfo{person}{Daniel Hsu}.} \bibinfo{year}{2011}\natexlab{}.
\newblock \showarticletitle{Sample complexity bounds for differentially private
  learning}. In \bibinfo{booktitle}{\emph{Proceedings of the 24th Annual
  Conference on Learning Theory}}. JMLR Workshop and Conference Proceedings,
  \bibinfo{pages}{155--186}.
\newblock


\bibitem[Chillotti et~al\mbox{.}(2020)]%
        {chillotti2020tfhe}
\bibfield{author}{\bibinfo{person}{Ilaria Chillotti}, \bibinfo{person}{Nicolas
  Gama}, \bibinfo{person}{Mariya Georgieva}, {and} \bibinfo{person}{Malika
  Izabach{\`e}ne}.} \bibinfo{year}{2020}\natexlab{}.
\newblock \showarticletitle{TFHE: fast fully homomorphic encryption over the
  torus}.
\newblock \bibinfo{journal}{\emph{Journal of Cryptology}} \bibinfo{volume}{33},
  \bibinfo{number}{1} (\bibinfo{year}{2020}), \bibinfo{pages}{34--91}.
\newblock


\bibitem[competition(2014)]%
        {purchase-dataset}
\bibfield{author}{\bibinfo{person}{Kaggle competition}.}
  \bibinfo{year}{2014}\natexlab{}.
\newblock \bibinfo{title}{Acquire Valued Shoppers Challenge}.
\newblock
\newblock
\urldef\tempurl%
\url{https://www.kaggle.com/c/acquire-valued-shoppers-challenge/overview}
\showURL{%
\tempurl}


\bibitem[Corrigan-Gibbs and Boneh(2017)]%
        {corrigan2017prio}
\bibfield{author}{\bibinfo{person}{Henry Corrigan-Gibbs} {and}
  \bibinfo{person}{Dan Boneh}.} \bibinfo{year}{2017}\natexlab{}.
\newblock \showarticletitle{Prio: Private, robust, and scalable computation of
  aggregate statistics}. In \bibinfo{booktitle}{\emph{14th $\{$USENIX$\}$
  Symposium on Networked Systems Design and Implementation ($\{$NSDI$\}$ 17)}}.
  \bibinfo{pages}{259--282}.
\newblock


\bibitem[Dong et~al\mbox{.}(2022)]%
        {dong2022gaussian}
\bibfield{author}{\bibinfo{person}{Jinshuo Dong}, \bibinfo{person}{Aaron Roth},
  {and} \bibinfo{person}{Weijie~J Su}.} \bibinfo{year}{2022}\natexlab{}.
\newblock \showarticletitle{Gaussian differential privacy}.
\newblock \bibinfo{journal}{\emph{Journal of the Royal Statistical Society
  Series B: Statistical Methodology}} \bibinfo{volume}{84}, \bibinfo{number}{1}
  (\bibinfo{year}{2022}), \bibinfo{pages}{3--37}.
\newblock


\bibitem[Dwork et~al\mbox{.}(2006)]%
        {dwork2006calibrating}
\bibfield{author}{\bibinfo{person}{Cynthia Dwork}, \bibinfo{person}{Frank
  McSherry}, \bibinfo{person}{Kobbi Nissim}, {and} \bibinfo{person}{Adam
  Smith}.} \bibinfo{year}{2006}\natexlab{}.
\newblock \showarticletitle{Calibrating noise to sensitivity in private data
  analysis}. In \bibinfo{booktitle}{\emph{Theory of Cryptography: Third Theory
  of Cryptography Conference, TCC 2006, New York, NY, USA, March 4-7, 2006.
  Proceedings 3}}. Springer, \bibinfo{pages}{265--284}.
\newblock


\bibitem[{Froelicher} et~al\mbox{.}(2020)]%
        {Froelicher2019}
\bibfield{author}{\bibinfo{person}{D. {Froelicher}}, \bibinfo{person}{J.~R.
  {Troncoso-Pastoriza}}, \bibinfo{person}{J.~S. {Sousa}}, {and}
  \bibinfo{person}{J. {Hubaux}}.} \bibinfo{year}{2020}\natexlab{}.
\newblock \showarticletitle{Drynx: Decentralized, Secure, Verifiable System for
  Statistical Queries and Machine Learning on Distributed Datasets}.
\newblock \bibinfo{journal}{\emph{IEEE Transactions on Information Forensics
  and Security}}  \bibinfo{volume}{15} (\bibinfo{year}{2020}),
  \bibinfo{pages}{3035--3050}.
\newblock


\bibitem[Halevi and Shoup(2020)]%
        {halevi2020helib}
\bibfield{author}{\bibinfo{person}{Shai Halevi} {and} \bibinfo{person}{Victor
  Shoup}.} \bibinfo{year}{2020}\natexlab{}.
\newblock \bibinfo{title}{Design and implementation of HElib: a homomorphic
  encryption library}.
\newblock \bibinfo{howpublished}{Cryptology ePrint Archive, Paper 2020/1481}.
\newblock


\bibitem[Hesamifard et~al\mbox{.}(2018)]%
        {hesamifard2018privacy}
\bibfield{author}{\bibinfo{person}{Ehsan Hesamifard}, \bibinfo{person}{Hassan
  Takabi}, \bibinfo{person}{Mehdi Ghasemi}, {and} \bibinfo{person}{Rebecca~N
  Wright}.} \bibinfo{year}{2018}\natexlab{}.
\newblock \showarticletitle{Privacy-preserving machine learning as a service.}
\newblock \bibinfo{journal}{\emph{Proc. Priv. Enhancing Technol.}}
  \bibinfo{volume}{2018}, \bibinfo{number}{3} (\bibinfo{year}{2018}),
  \bibinfo{pages}{123--142}.
\newblock


\bibitem[Joye(2022)]%
        {joye2021guide}
\bibfield{author}{\bibinfo{person}{Marc Joye}.}
  \bibinfo{year}{2022}\natexlab{}.
\newblock \showarticletitle{SoK: Fully homomorphic encryption over the
  [discretized] torus}.
\newblock \bibinfo{journal}{\emph{IACR Transactions on Cryptographic Hardware
  and Embedded Systems}} (\bibinfo{year}{2022}), \bibinfo{pages}{661--692}.
\newblock


\bibitem[Keller and Sun(2022)]%
        {keller2022secure}
\bibfield{author}{\bibinfo{person}{Marcel Keller} {and} \bibinfo{person}{Ke
  Sun}.} \bibinfo{year}{2022}\natexlab{}.
\newblock \showarticletitle{Secure quantized training for deep learning}. In
  \bibinfo{booktitle}{\emph{International Conference on Machine Learning}}.
  PMLR, \bibinfo{pages}{10912--10938}.
\newblock


\bibitem[Kelly et~al\mbox{.}(2022)]%
        {uci-datasets}
\bibfield{author}{\bibinfo{person}{Markelle Kelly}, \bibinfo{person}{Rachel
  Longjohn}, {and} \bibinfo{person}{Kolby Nottingham}.}
  \bibinfo{year}{2022}\natexlab{}.
\newblock \bibinfo{title}{The UCI Machine Learning Repository}.
\newblock
\newblock
\urldef\tempurl%
\url{https://archive.ics.uci.edu}
\showURL{%
\tempurl}


\bibitem[Krizhevsky et~al\mbox{.}(2009)]%
        {cifar-datasets}
\bibfield{author}{\bibinfo{person}{Alex Krizhevsky} {et~al\mbox{.}}}
  \bibinfo{year}{2009}\natexlab{}.
\newblock \showarticletitle{Learning multiple layers of features from tiny
  images}.
\newblock  (\bibinfo{year}{2009}).
\newblock


\bibitem[Lou et~al\mbox{.}(2020)]%
        {lou2020glyph}
\bibfield{author}{\bibinfo{person}{Qian Lou}, \bibinfo{person}{Bo Feng},
  \bibinfo{person}{Geoffrey Charles~Fox}, {and} \bibinfo{person}{Lei Jiang}.}
  \bibinfo{year}{2020}\natexlab{}.
\newblock \showarticletitle{Glyph: Fast and accurately training deep neural
  networks on encrypted data}.
\newblock \bibinfo{journal}{\emph{Advances in Neural Information Processing
  Systems}}  \bibinfo{volume}{33} (\bibinfo{year}{2020}),
  \bibinfo{pages}{9193--9202}.
\newblock


\bibitem[Lu et~al\mbox{.}(2022)]%
        {lu2022differentially}
\bibfield{author}{\bibinfo{person}{Zhigang Lu}, \bibinfo{person}{Hassan~Jameel
  Asghar}, \bibinfo{person}{Mohamed~Ali Kaafar}, \bibinfo{person}{Darren Webb},
  {and} \bibinfo{person}{Peter Dickinson}.} \bibinfo{year}{2022}\natexlab{}.
\newblock \showarticletitle{A differentially private framework for deep
  learning with convexified loss functions}.
\newblock \bibinfo{journal}{\emph{IEEE Transactions on Information Forensics
  and Security}}  \bibinfo{volume}{17} (\bibinfo{year}{2022}),
  \bibinfo{pages}{2151--2165}.
\newblock


\bibitem[López~Lobo(2020)]%
        {concept-drift}
\bibfield{author}{\bibinfo{person}{Jesús López~Lobo}.}
  \bibinfo{year}{2020}\natexlab{}.
\newblock \bibinfo{title}{{Synthetic datasets for concept drift detection
  purposes}}.
\newblock
\newblock
\urldef\tempurl%
\url{https://doi.org/10.7910/DVN/5OWRGB}
\showDOI{\tempurl}


\bibitem[Mohassel and Zhang(2017)]%
        {mohassel2017secureml}
\bibfield{author}{\bibinfo{person}{Payman Mohassel} {and}
  \bibinfo{person}{Yupeng Zhang}.} \bibinfo{year}{2017}\natexlab{}.
\newblock \showarticletitle{Secureml: A system for scalable privacy-preserving
  machine learning}. In \bibinfo{booktitle}{\emph{2017 IEEE symposium on
  security and privacy (SP)}}. IEEE, \bibinfo{pages}{19--38}.
\newblock


\bibitem[Morais et~al\mbox{.}(2019)]%
        {morais2019zkrp}
\bibfield{author}{\bibinfo{person}{Eduardo Morais}, \bibinfo{person}{Tommy
  Koens}, \bibinfo{person}{Cees Van~Wijk}, {and} \bibinfo{person}{Aleksei
  Koren}.} \bibinfo{year}{2019}\natexlab{}.
\newblock \showarticletitle{A survey on zero knowledge range proofs and
  applications}.
\newblock \bibinfo{journal}{\emph{SN Applied Sciences}}  \bibinfo{volume}{1}
  (\bibinfo{year}{2019}), \bibinfo{pages}{1--17}.
\newblock


\bibitem[Nandakumar et~al\mbox{.}(2019)]%
        {nandakumar2019towards}
\bibfield{author}{\bibinfo{person}{Karthik Nandakumar}, \bibinfo{person}{Nalini
  Ratha}, \bibinfo{person}{Sharath Pankanti}, {and} \bibinfo{person}{Shai
  Halevi}.} \bibinfo{year}{2019}\natexlab{}.
\newblock \showarticletitle{Towards deep neural network training on encrypted
  data}. In \bibinfo{booktitle}{\emph{Proceedings of the IEEE/CVF Conference on
  Computer Vision and Pattern Recognition Workshops}}. \bibinfo{pages}{0--0}.
\newblock


\bibitem[Shalev-Shwartz and Ben-David(2014)]%
        {shai-shai-book}
\bibfield{author}{\bibinfo{person}{Shai Shalev-Shwartz} {and}
  \bibinfo{person}{Shai Ben-David}.} \bibinfo{year}{2014}\natexlab{}.
\newblock \bibinfo{booktitle}{\emph{Understanding machine learning: From theory
  to algorithms}}.
\newblock \bibinfo{publisher}{Cambridge university press}.
\newblock


\bibitem[Shokri et~al\mbox{.}(2017)]%
        {shokri2017membership}
\bibfield{author}{\bibinfo{person}{Reza Shokri}, \bibinfo{person}{Marco
  Stronati}, \bibinfo{person}{Congzheng Song}, {and} \bibinfo{person}{Vitaly
  Shmatikov}.} \bibinfo{year}{2017}\natexlab{}.
\newblock \showarticletitle{Membership inference attacks against machine
  learning models}. In \bibinfo{booktitle}{\emph{2017 IEEE symposium on
  security and privacy (SP)}}. IEEE, \bibinfo{pages}{3--18}.
\newblock


\bibitem[Smith et~al\mbox{.}(2022)]%
        {smith2022making}
\bibfield{author}{\bibinfo{person}{Josh Smith}, \bibinfo{person}{Hassan~Jameel
  Asghar}, \bibinfo{person}{Gianpaolo Gioiosa}, \bibinfo{person}{Sirine
  Mrabet}, \bibinfo{person}{Serge Gaspers}, {and} \bibinfo{person}{Paul
  Tyler}.} \bibinfo{year}{2022}\natexlab{}.
\newblock \showarticletitle{Making the Most of Parallel Composition in
  Differential Privacy}.
\newblock \bibinfo{journal}{\emph{Proceedings on Privacy Enhancing
  Technologies}}  \bibinfo{volume}{1} (\bibinfo{year}{2022}),
  \bibinfo{pages}{253--273}.
\newblock


\bibitem[Thomas(2017)]%
        {lief}
\bibfield{author}{\bibinfo{person}{Romain Thomas}.}
  \bibinfo{year}{2017}\natexlab{}.
\newblock \bibinfo{title}{LIEF - Library to Instrument Executable Formats}.
\newblock \bibinfo{howpublished}{https://lief.quarkslab.com/}.
\newblock


\bibitem[Widmer and Kubat(1996)]%
        {widmer1996conceptdrift}
\bibfield{author}{\bibinfo{person}{Gerhard Widmer} {and}
  \bibinfo{person}{Miroslav Kubat}.} \bibinfo{year}{1996}\natexlab{}.
\newblock \showarticletitle{Learning in the presence of concept drift and
  hidden contexts}.
\newblock \bibinfo{journal}{\emph{Machine learning}}  \bibinfo{volume}{23}
  (\bibinfo{year}{1996}), \bibinfo{pages}{69--101}.
\newblock


\bibitem[Xu et~al\mbox{.}(2022)]%
        {xu2021nn}
\bibfield{author}{\bibinfo{person}{Runhua Xu}, \bibinfo{person}{James Joshi},
  {and} \bibinfo{person}{Chao Li}.} \bibinfo{year}{2022}\natexlab{}.
\newblock \showarticletitle{NN-EMD: Efficiently Training Neural Networks Using
  Encrypted Multi-Sourced Datasets}.
\newblock \bibinfo{journal}{\emph{IEEE Transactions on Dependable and Secure
  Computing}} \bibinfo{volume}{19}, \bibinfo{number}{4} (\bibinfo{year}{2022}),
  \bibinfo{pages}{2807--2820}.
\newblock
\urldef\tempurl%
\url{https://doi.org/10.1109/TDSC.2021.3074439}
\showDOI{\tempurl}


\bibitem[Xu et~al\mbox{.}(2019)]%
        {xu2019cryptonn}
\bibfield{author}{\bibinfo{person}{Runhua Xu}, \bibinfo{person}{James~BD
  Joshi}, {and} \bibinfo{person}{Chao Li}.} \bibinfo{year}{2019}\natexlab{}.
\newblock \showarticletitle{Cryptonn: Training neural networks over encrypted
  data}. In \bibinfo{booktitle}{\emph{2019 IEEE 39th International Conference
  on Distributed Computing Systems (ICDCS)}}. IEEE,
  \bibinfo{pages}{1199--1209}.
\newblock


\bibitem[Yuan et~al\mbox{.}(2021)]%
        {yuan2021label}
\bibfield{author}{\bibinfo{person}{Sen Yuan}, \bibinfo{person}{Milan Shen},
  \bibinfo{person}{Ilya Mironov}, {and} \bibinfo{person}{Anderson Nascimento}.}
  \bibinfo{year}{2021}\natexlab{}.
\newblock \showarticletitle{Label Private Deep Learning Training based on
  Secure Multiparty Computation and Differential Privacy}. In
  \bibinfo{booktitle}{\emph{NeurIPS 2021 Workshop Privacy in Machine
  Learning}}.
\newblock


\bibitem[Zahoora et~al\mbox{.}(2022)]%
        {zahoora2022ransomware}
\bibfield{author}{\bibinfo{person}{Umme Zahoora}, \bibinfo{person}{Asifullah
  Khan}, \bibinfo{person}{Muttukrishnan Rajarajan},
  \bibinfo{person}{Saddam~Hussain Khan}, \bibinfo{person}{Muhammad Asam}, {and}
  \bibinfo{person}{Tauseef Jamal}.} \bibinfo{year}{2022}\natexlab{}.
\newblock \showarticletitle{Ransomware detection using deep learning based
  unsupervised feature extraction and a cost sensitive Pareto Ensemble
  classifier}.
\newblock \bibinfo{journal}{\emph{Scientific reports}} \bibinfo{volume}{12},
  \bibinfo{number}{1} (\bibinfo{year}{2022}), \bibinfo{pages}{15647}.
\newblock


\bibitem[Zama(2022)]%
        {concrete}
\bibfield{author}{\bibinfo{person}{Zama}.} \bibinfo{year}{2022}\natexlab{}.
\newblock \bibinfo{title}{{Concrete: TFHE Compiler that converts python
  programs into FHE equivalent}}.
\newblock
\newblock
\newblock
\shownote{\url{https://github.com/zama-ai/concrete}}.


\end{thebibliography}

\appendix


\section{Proofs of Theorems from Section~\ref{sec:accuracy-gain}}
\label{app:knowledge-proofs}

\subsection*{Proof of Theorem~\ref{lem:half-loss-balanced}}
\begin{proof}
\revise{Let the samples in $D_{\text{hold}}$ be $(x_i, y_i)$ for $i \in [m]$. Let $\mathbbm{1}_{[M(x_i) \neq y_i]}$ denote the indicator random variable which is 1 if $M(x_i) \neq y_i$ for each $i \in [m]$. Then through the linearity of expectation:
\begin{align*}
    \mathbb{E}_g[L_\text{hold}(M)] &= \frac{1}{m} \sum_{i=1}^m \mathbb{E}_g [\mathbbm{1}_{[M(x_i) \neq y_i]}]\\
    &= \frac{1}{m} \sum_{i=1}^m \Pr_g[M(x_i) \neq y_i]\\
    &= \Pr_{g, (x,y) \sim D_\text{hold}}[M(x) \neq y],
\end{align*}
where the subscript $g$ indicates that probability is taken over the randomness in $g$. The last result follows from the law of total probability. Recall that the notation $(x, y) \sim D_\text{hold}$ means the sample is chosen uniformly at random from $D_\text{hold}$. Dropping subscripts
}
consider the probability $\Pr [M(x) \neq y]$.
We have
\begin{align*}
    \Pr [M(x) \neq y] &= \Pr [M(x) \neq y \mid y = 1 ] \Pr [y = 1] \\
    &+ \Pr [M(x) \neq y \mid y = 0] \Pr [y = 0] \\
    &= \Pr [M(x) = 0 \mid y = 1 ] \Pr [y = 1] \\
    &+ \Pr [M(x) = 1 \mid y = 0] \Pr [y = 0]
\end{align*}
Now, the learning algorithm $\mathcal{A}$'s input, i.e., $D$, remains unchanged whether $y$, i.e., the label of $x$ in $D_\text{hold}$,  is equal to 0 and 1. This is because $D$ is labelled by $g$ which is independent of the true label $y$ (Eq.~\ref{eq:lab-func}). Therefore, 
\[
\Pr [M(x) = 0 \mid y = 1 ] = \Pr [M(x) = 0 ],
\]
and 
\[
\Pr [M(x) = 1 \mid y = 0 ] = \Pr [M(x) = 1 ].
\]
We get:
\begin{align*}
     \Pr [M(x) \neq y] &= \Pr [M(x) = 0] \Pr [y = 1] \\
    &+ \Pr [M(x) = 1] \Pr [y = 0]\\
    &= \Pr [M(x) = 0] \frac{1}{2} + \Pr [M(x) = 1] \frac{1}{2}\\
    &= \frac{1}{2}(\Pr [M(x) = 0] + \Pr [M(x) = 1] ) = \frac{1}{2}
\end{align*}
Therefore $\mathbb{E}[{L}_{\text{hold}}(M)] = \Pr_g [M(x) \neq y] = \frac{1}{2}$, where $(x, y) \sim D_\text{hold}$.  
\end{proof}

\subsection*{Proof of Theorem~\ref{theo:m2-less-than-m1}}
\begin{proof}
   Since $D_\text{hold}$ is balanced, for $(x, y) \sim D_\text{hold}$, and following the proof of Theorem~\ref{lem:half-loss-balanced}, we get
\begin{align*}
    \mathbb{E}_g[L_\text{hold}(M_2)] &= 
\Pr_{g, (x,y) \sim D_\text{hold}} [M_2(x) \neq y] \\
&= \Pr_g [M_2(x) = 0 \mid y = 1 ] \Pr [y = 1] \\
    &+ \Pr_g [M_2(x) = 1 \mid y = 0] \Pr [y = 0] \\
    &= \frac{1}{2}(\Pr_g [M_2(x) = 0 \mid y = 1 ] \\
    &+ \Pr_g [M_2(x) = 1 \mid y = 0] ) \\
    &\geq \frac{1}{2}(\Pr [M_1(x) = 0 \mid y = 1 ] \\
    &+ \Pr [M_1(x) = 1 \mid y = 0] ) \\
    &= \Pr [M_1(x) \neq y] = L_{\text{hold}}(M_1),
\end{align*}
where the inequality follows from condition~\eqref{eq:oea} in the theorem statement, and we have dropped the subscript $(x, y) \sim D_\text{hold}$ from the second equality onwards for better readability. 
\end{proof}

\revise{
\subsection*{Proof of Theorem~\ref{theo:hoeffding}}
\begin{proof}
Let 
\begin{equation}
\label{eq:th3:1}
\delta \leq L_\text{hold}(M_1) - L_\text{hold}(M_2) = (1 - L_\text{hold}(M_2)) - (1 - L_\text{hold}(M_1)).
\end{equation}
From Theorem~\ref{theo:m2-less-than-m1}, we know that 
\begin{align}
 \mathbb{E}_g[L_\text{hold}(M_2)] &\geq  L_\text{hold}(M_1) \nonumber\\
 1 - \mathbb{E}_g[L_\text{hold}(M_2)] &\leq 1 - L_\text{hold}(M_1), \nonumber
\end{align}
Giving us:
\begin{align}
 & (1 - L_\text{hold}(M_2)) - (1 - \mathbb{E}_g[L_\text{hold}(M_2)]) \nonumber \\
 &\geq (1 - L_\text{hold}(M_2)) - (1 - L_\text{hold}(M_1)) \geq \delta \label{eq:th3:2}
\end{align}
Let us call the first term $\mathsf{A}_1$ and the second $\mathsf{A}_2$ in the above three-term inequality. Then 
\begin{equation}
    \Pr[\mathsf{A}_1 \geq \delta] \leq \Pr[\mathsf{A}_2 \geq \delta], \label{eq:th3:3}
\end{equation}
since whenever the event $\mathsf{A}_2 \geq \delta$ occurs, the event $\mathsf{A}_1 \geq \delta$ is guaranteed to occur due to Eq.~\eqref{eq:th3:2}. Now, consider
\begin{align}
    1 - L_\text{hold}(M_2) &= 1 - \frac{\sum_{i = 1}^m \mathbbm{1}_{[M_2(x_i) \neq y_i]}}{m} \nonumber\\
    &= \frac{\sum_{i = 1}^m \left(1 - \mathbbm{1}_{[M_2(x_i) \neq y_i]}\right)}{m} \label{eq:th3:4},
\end{align}
where $\mathbbm{1}_{[M_2(x_i) \neq y_i]}$ is the indicator function which is 1 if the $i$th sample in $D_\text{hold}$, i.e., $(x_i, y_i)$ is mislabelled by $M_2$. Consider also:
\begin{align}
    1 - \mathbb{E}_g[L_\text{hold}(M_2)] &= 1 - \frac{\sum_{i = 1}^m \mathbb{E}_g[\mathbbm{1}_{[M_2(x_i) \neq y_i]}]}{m} \nonumber\\
    &= \frac{\sum_{i = 1}^m \mathbb{E}_g \left[1 - \mathbbm{1}_{[M_2(x_i) \neq y_i]}\right]}{m} \label{eq:th3:5},
\end{align}
Thus, $\mathsf{A}_1$ in Eq.~\eqref{eq:th3:3} includes the sum of $m$ random variables minus the sum of expected values of these $m$ random variables. Each of these variables are between 0 and 1, being indicator random variables. Therefore, we can invoke Hoeffding's inequality to conclude, combining Eqs~\ref{eq:th3:1}, \ref{eq:th3:2}, \ref{eq:th3:3}, \ref{eq:th3:4} and \ref{eq:th3:5} that:
\begin{align*}
    &\Pr_g [L_\text{hold}(M_1) - L_\text{hold}(M_2) 
    \geq \delta] \\
    &= \Pr_g [ (1 - L_\text{hold}(M_2)) - (1 - L_\text{hold}(M_1)) \geq \delta 
 ]\\
 & \leq \Pr_g [ (1 - L_\text{hold}(M_2)) - (1 - \mathbb{E}_g[L_\text{hold}(M_2)])] \\
 &\leq \exp(-2m \delta^2).
\end{align*}
\end{proof}
}

\revise{
\subsection*{When Does Condition~\eqref{eq:oea} Hold?}
The proofs of Theorems~\ref{theo:m2-less-than-m1} and \ref{theo:hoeffding} rely on condition~\eqref{eq:oea} to hold. The condition states that with the model and datasets as defined before, 
if $(x, y) \sim D_\text{hold}$ then
\begin{align*}
 &\Pr_g [M_2(x) = 0 \mid y = 1 ] + \Pr_g [M_2(x) = 1 \mid y = 0] \\
 &\geq \Pr [M_1(x) = 0 \mid y = 1 ] + \Pr [M_1(x) = 1 \mid y = 0],
\end{align*}
where the probabilities in the left hand side of the inequality are also over the random choices of $g$. Notice that this does not necessarily mean that one-sided errors or any particular $g$ may be worse off, i.e., we may have for a particular $g$ that $\Pr [M_2(x) = 0 \mid y = 1 ]  \leq \Pr [M_1(x) = 0 \mid y = 1 ]$, since $M_2$ may assign the label 1 to every sample.
}

\revise{The condition is not always satisfied. For example, the condition does not hold if the labels of $D_1$ are out of distribution. For instance, assume that almost all samples in $D_1$ have label 0. Further, assume that all the errors made by $M_1$ are on label 1. This is likely to be the case, as the model is trained overwhelmingly on samples with label 0. In this case, if $D_2$ has all samples labeled 1 (i.e., the oblivious labeling function $g(x) = 1$, for all $x$), then $M_2$ may have decreased error on samples with label 1, and therefore the overall error will be reduced. 
}

\revise{
On the other hand, if $D_1$ follows the original distribution $\mathcal{D}$, then since $D_2$ is labelled according to $g$, it does not increase $\mathcal{A}$'s information about the original conditional distribution $\mathcal{D}_{y | x}$. Therefore, the overall error of $M_2$ on $D_\text{hold}$, which follows the original conditional distribution $\mathcal{D}_{y | x}$, can at best be greater than or equal to that of $M_1$ on $D_\text{hold}$ over all possible random choices of the oblivious functions $g$, as both are outputs of the same learning algorithm $\mathcal{A}$. We postulate that in this case, the above condition holds. 
}

\revise{
To test this, we used the purchase dataset~\footnote{See \url{https://www.kaggle.com/c/acquire-valued-shoppers-challenge/data}.} clustered into two classes. We used the first 200,000 samples of the dataset. The dataset is relatively balanced with slightly more than 65\% of the samples having label 1. We reserved the first 100,000 rows for $D_1$. The remaining 100,000 rows were reserved for $D_2$ and $D_\text{hold}$. For $D_\text{hold}$, we took a 20\% sample from these 100,000 samples, and discarded samples so that we have 2,000 samples in $D_\text{hold}$ with exactly equal distribution of the two labels. Thus, 80,000 samples were reserved for $D_2$. We trained the Random Forest (RF) classifier from \texttt{scikit-learn} with default settings on $D_1$, and tested the accuracy on $D_\text{hold}$. The average accuracy over 100 runs was $\approx 0.90$. We then restricted $D_1$ to the first 2,000 samples to simulate lack of samples. We trained the RF classifier again on the updated $D_1$, and obtained average accuracy of $\approx 0.87$. Thus, there is a gap in the two accuracies, and hence room for improvement. We also trained the RF classifier on $D_2$. For a realistic scenario, we took $D_2$ to be the first 2,000, 3,000 and 5,000 samples (from the total of 80,000 available samples). This is a realistic setting as increasing the size of the combined dataset is likely to increase accuracy. The average accuracy of the RF classifier on $D_2$ alone was $\approx 0.89$, and on $D_1 \cup D_2$ was $\approx 0.89$, both over 100 runs. Thus, the collaboration results in a better model if the two datasets are joined. 
}

\revise{
Now to check if Theorem~\ref{theo:hoeffding} holds, we chose $\delta = 0.01$ and $\delta = 0.02$, and selected different (balanced) sizes of $D_\text{hold}$ upto a maximum size of 2,000. The oblivious labeling function $g_q$ was chosen such that it assigns the label $1$ to a given sample with probability $1-q$, and 0 otherwise, where $q \in \{0, 0.1, 0.2, \ldots, 1\}$. This constitutes a subset of all possible oblivious labeling functions. With these parameters, for each size of $D_\text{hold}$ in the above set, and for each value of $q$, we trained the RF classifier on $D_1 \cup D'_2$ where $D'_2$ had the same samples (features) as $D_2$, except that the labels were labeled by $g_q$. We checked if the resulting accuracy was greater than that obtained from the model trained on $D_1$ plus $\delta$. This process was repeated 100 times for each configuration, and the average number of times the resulting accuracy was greater than this quantity was noted as the estimated probability of Theorem~\ref{theo:hoeffding}. 
}

\begin{figure*}[!ht]
    \centering
    \captionsetup{justification=centering}
    \subfloat{
    \includegraphics[width=0.32\textwidth]{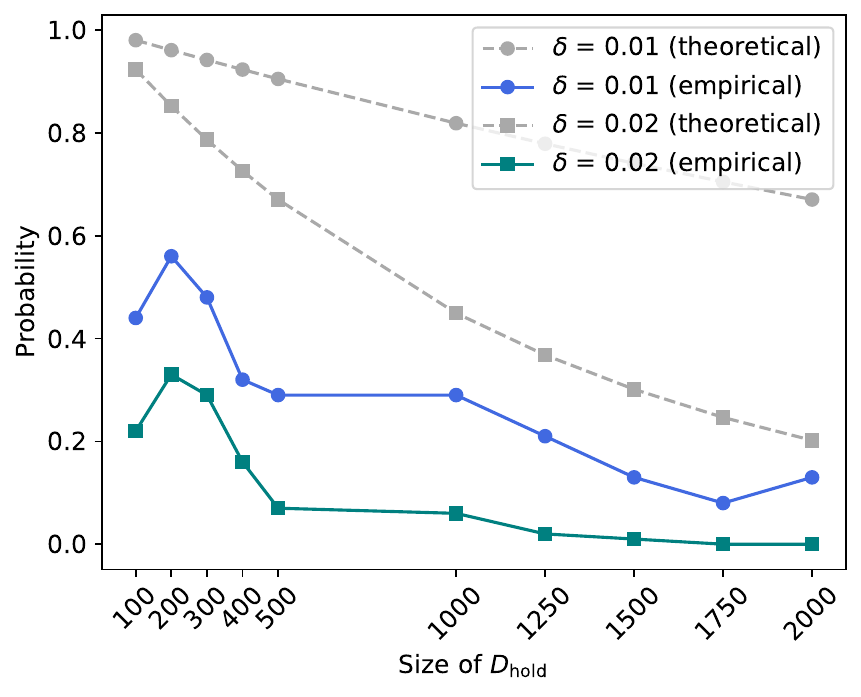}
    \label{subfig:ct-2000}
    }
   \hfill
    \subfloat{
    \includegraphics[width=0.32\textwidth]{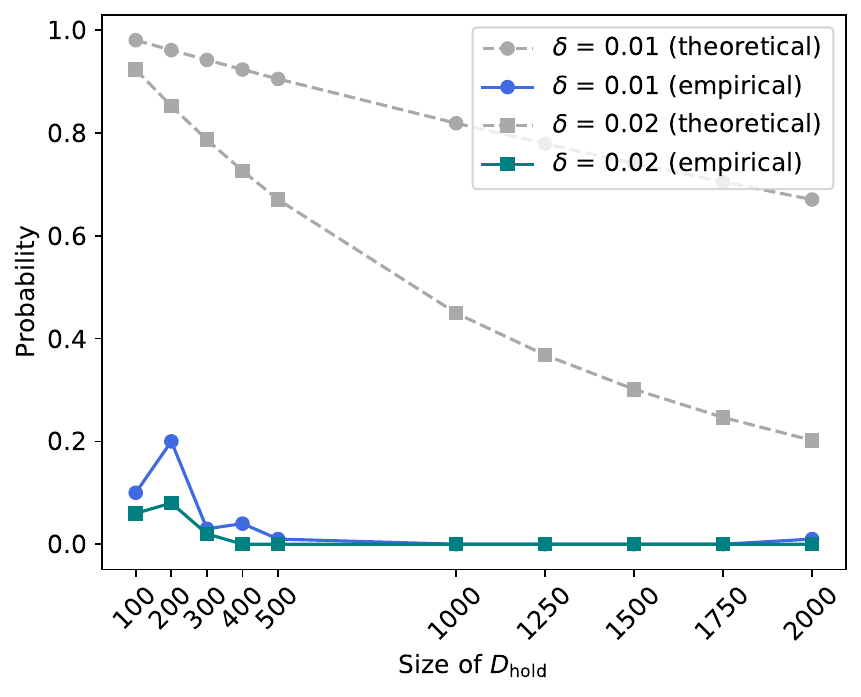}
    \label{subfig:ct-3000}
    }
    \hfill
    \subfloat{
    \includegraphics[width=0.32\textwidth]{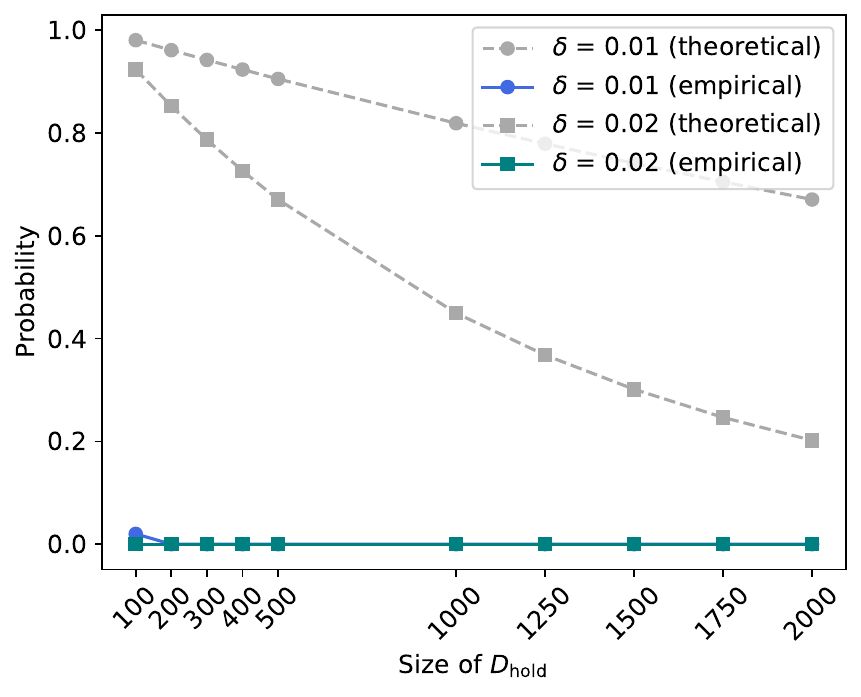}
    }
    \caption{\revise{The number of times $L_\text{hold}(M_2) \leq L_\text{hold}(M_1) + \delta$, where $M_1$ is trained on $D_1$ and $M_2$ on $D_1 \cup D'_2$, and $D'_2$ labelled through an oblivious function. The size of $D'_2$ in the leftmost image is 2,000, for the middle image it is 3,000 and for the rightmost image it is 5,000. The empirical probabilities are far below the theoretical bound from Theorem~\ref{theo:hoeffding} as the size of $D'_2$ grows.}}
    \label{fig:check-theorem}
\end{figure*}

\revise{
Figure~\ref{fig:check-theorem} shows the results of this experiment. For each size of $D_\text{hold}$, we pick the highest probability against all possible values of $q$ in $g_q$. As we can see, the empirical probabilities are lower than the estimated bound from Theorem~\ref{theo:hoeffding}. Furthermore, with increasing size of $D'_2$, the estimated probabilities are significantly less than the theoretical bound. Thus, in this case the theorem holds. A comprehensive analysis of when the condition holds and whether it is true for other classifiers and datasets requires significant effort. We leave it as an interesting open question.
}

\revise{
\section{Differential Privacy Violation with Integer Encoded Gaussian Noise}
\label{app:dp-violate}
In this section, we show that multiplying encoded sensitivity with encoded noise violates differential privacy. This is the main reason why we need the list of allowable sensitivities. This vulnerability also exists in the work from Yuan et al~\cite{yuan2021label}. For ease of exposition, consider the one dimensional case, which can be straighforwardly extended to higher dimensions. Suppose we have a function $\ell(x)$ whose sensitivity is $s$ over neighbouring datasets. We therefore add  Gaussian noise $s\mathcal{N}(0, \sigma^2) = \mathcal{N}(0, (s \sigma)^2)$ to $\ell(x)$ to make it differentially private, where for completeness assume that $\sigma = 1/\epsilon$. However, in our case, and also in~\cite{yuan2021label}, the sensitivity and the noise are computed by two different parties. We therefore first need to encrypt them and then homomorphically multiply. Encryption implies that these quantities need to be integer encoded. But in general the encoded versions do not satisfy the above relation, i.e.,
\[
 \lfloor rs \rfloor \cdot \lfloor  \mathcal{N}(0, \sigma^2)\rfloor  \neq \lfloor rs\mathcal{N}(0, \sigma)^2\rfloor = \lfloor \mathcal{N}(0, (rs\sigma)^2)\rfloor,
\]
where $r$ is the precision used in encoding. For instance, consider $r = 10^6, s = 1.2$, and $\mathcal{N}(0, \sigma^2) = 1.1$. Then 
\[
\lfloor \mathcal{N}(0, (rs\sigma)^2) \rfloor = \lfloor rs\mathcal{N}(0, \sigma^2) \rfloor = \lfloor 10^6 \times 1.2 \times 1.1 \rfloor = 1.32 \times 10^6,
\]
whereas 
\[
\lfloor rs \rfloor \cdot \lfloor  \mathcal{N}(0, \sigma^2)\rfloor = \lfloor 10^6 \times 1.2 \rfloor \cdot \lfloor 1.1 \rfloor   = 1.2 \times 10^6.
\]
Note that the authors in~\cite{yuan2021label} mention that to integer encode their scheme one can use the encoding scheme used in~\cite{xu2019cryptonn}, which is the same as ours. The fact that the encoded noise does not follow the said property means that the resulting mechanism is no longer differentially private. To see this, again for simplicity consider one dimension and further assume that we have one sample in the batch and the sensitivity is $s = 2$, and therefore $rs = 2r$. Then the encoded noise is a multiple of $2r$, regardless of the actual noise value and the derivative of the loss function. Now, if the initial label is $y = 0$, the quantity with noise remains a multiple of $2r$. However, if the label is $y = 1$, then the quantity with noise is $r$ plus a multiple of $2r$. Thus after unblinding (step 10), $P_1$ will be able to determine the label as it knows the sensitivity.\footnote{In Protocol 2 from Yuan et al.~\cite{yuan2021label}, Alice will be able to determine the label as she computed the sensitivity. Note that in their protocol it is the quantity $p-y$ which is encoded, where $p$ is the predicted confidence (probability) associated with label $y$. The reasoning holds by assume $p = 1$ for the input sample.} This reasoning can be extended to the entire mini-batch by assuming the neighbouring datasets to only differ in the label $y$.  
}

\section{Computing Gradients via the Chain Rule}
\label{app:grad}
From Equation~\eqref{eq:gradients}, we are interested in computing the loss through the samples in a batch $B$ that belong to the dataset $D_2$. Overloading notation, we still use $B$ to denote the samples belonging to $D_2$. The algorithm used to minimizing the loss is the stochastic gradient descent algorithm using backpropagation. This inolves calculating the gradient $\nabla L_B(\mathbf{w})$. As noted in~\cite{yuan2021label}, if we are using the backpropagation algorithm, we only need to be concerned about the gradients corresponding to the last layer. Again, to simplify notation, we denote the vector of weights in the \emph{last layer} by $\mathbf{w}$. 

Let $s = (\mathbf{x}, \mathbf{y})$ be a sample in the batch $B$. Let $K$ denote the number of classes. We assume that $\mathbf{y}$ is one-hot encoded, meaning that only one of its element is 1, and the rest are zero. We denote the index of this element by $c$ (which of course is different for different samples). The output from the neural network is the softmax output $\mathbf{p}$ of size $K$, where:
\[
p_i = \frac{e^{z_i}}{\sum_j e^{z_j}},
\]
where $\mathbf{z} = (z_1, \ldots, z_K)$ is the output before the softmax layer. Consider the cross-entropy loss. Under this loss, we have:
\begin{equation}
\label{eq:ce-loss}
    L_s(\mathbf{w}) = - \sum_{i = 1}^K y_i \ln p_i = - \ln p_c,
\end{equation}
since $\mathbf{y}$ is one-hot encoded. Now, we have:
\begin{equation}
\label{eq:app:overall-L}
\nabla L_B(\mathbf{w}) = \frac{1}{|B|} \sum_{s \in B} \nabla L_s(\mathbf{w})
\end{equation}

Let us, therefore, focus on the gradient of per-sample loss. By the chain rule, we have:
\begin{equation}
\label{eq:app:chain-rule-L}
 \nabla L_s(\mathbf{w}) = \frac{\partial L}{\partial \mathbf{w}} = \frac{\partial L}{\partial \mathbf{z}} \frac{\partial \mathbf{z}}{\partial \mathbf{w}} 
\end{equation}
Consider the $i$th element of $\frac{\partial L}{\partial \mathbf{z}}$:
\begin{align}
\label{eq:app:loss-zi}
    \frac{\partial L}{\partial z_i} &= \frac{\partial }{\partial z_i} \left( - \ln p_c \right) = - \frac{1}{p_c} \frac{\partial p_c}{\partial z_i}.
\end{align}
Solving for the case when $i = c$, the above becomes $p_c - 1$, whereas for the case $i \neq c$, we get $p_i$. In both cases, we get:
\[
\frac{\partial L}{\partial z_i} = p_i - \mathbf{1}_c(i) = p_i - y_i
\]
Plugging this into Equation~\eqref{eq:app:chain-rule-L}, we get:
\begin{equation}
\label{eq:app:sep-L}
 \nabla L_s(\mathbf{w}) = \sum_{i = 1}^K (p_i - \mathbf{1}_c(i) )\frac{\partial z_i}{\partial \mathbf{w}} 
\end{equation}
Plugging this into Equation~\eqref{eq:app:overall-L}, we get:
\begin{align}
    \nabla L_B(\mathbf{w}) &= \frac{1}{|B|} \sum_{s \in B} \sum_{i = 1}^K (p_i(s) - \mathbf{1}_{c(s)}(i) )\frac{\partial z_i(s)}{\partial \mathbf{w}} \nonumber\\
    &= \frac{1}{|B|} \sum_{s \in B} \sum_{i = 1}^K p_i(s) \frac{\partial z_i(s)}{\partial \mathbf{w}} \nonumber\\
    &- \frac{1}{|B|} \sum_{s \in B} \sum_{i = 1}^K \mathbf{1}_{c(s)}(i)\frac{\partial z_i(s)}{\partial \mathbf{w}}, \label{eq:app:two-terms}
\end{align}
where $c(s), p_i(s)$ and $z_i(s)$ denote the fact that these quantities depend on the sample $s$. 

\section{Sensitivity of the Gradient of the Loss Function}
\label{app:sense}
We now compute the $\ell_2$-sensitivity of the loss function. Let $D'$ and $D''$ be neighbouring datasets, where all but one of the samples have the labels changed to another label. Let $s'$ and $s''$ represent the differing samples. Let $B'$ and $B''$ be the batches drawn in the two datasets. Note that $|B'| = |B''|$ and we denote this by $|B|$. Then, the $\ell_2$-sensitivity of the loss function, when no encoding is employed, is:

\begin{equation}
\begin{aligned}
    &\lVert \nabla L_{B'}(\mathbf{w}) - \nabla L_{B''}(\mathbf{w}) \rVert_2 \\
    &= \frac{1}{|B|} \left\lVert \sum_{s \in B'} \nabla L_s(\mathbf{w}) - \sum_{s \in B''} \nabla L_s(\mathbf{w}) \right\rVert_2 \nonumber\\
    &= \frac{1}{|B|} \left\lVert \nabla L_{s'}(\mathbf{w}) - \nabla L_{s''}(\mathbf{w}) \right\rVert_2 \nonumber \\
    &= \frac{1}{|B|} \left\lVert \sum_{i = 1}^K (p_i - \mathbf{1}_{c(s')}(i) )\frac{\partial z_i}{\partial \mathbf{w}} \right. \\ &- \left. \sum_{i = 1}^K (p_i - \mathbf{1}_{c(s'')}(i) )\frac{\partial z_i}{\partial \mathbf{w}}  \right\rVert_2 \nonumber \\
    &= \frac{1}{|B|} \left\lVert \sum_{i = 1}^K ( \mathbf{1}_{c(s'')}(i) - \mathbf{1}_{c(s')}(i) )\frac{\partial z_i}{\partial \mathbf{w}} \right\rVert_2 \nonumber\\
    &\leq \frac{1}{|B|} \sum_{i = 1}^K \left\lVert  ( \mathbf{1}_{c(s'')}(i) - \mathbf{1}_{c(s')}(i) )\frac{\partial z_i}{\partial \mathbf{w}} \right\rVert_2 \nonumber\\
    &= \frac{1}{|B|} \sum_{i = 1}^K \left\lvert  \mathbf{1}_{c(s'')}(i) - \mathbf{1}_{c(s')}(i) \right\rvert \left\lVert \frac{\partial z_i}{\partial \mathbf{w}} \right\rVert_2 \nonumber\\
    &\leq\frac{1}{|B|} \left(\sum_{i = 1}^K \left\lvert  \mathbf{1}_{c(s'')}(i) - \mathbf{1}_{c(s')}(i) \right\rvert \right) \max_i \left\lVert    \frac{\partial z_i}{\partial \mathbf{w}} \right\rVert_2 \nonumber\\
    &= \frac{2}{|B|} \max_i \left\lVert    \frac{\partial z_i}{\partial \mathbf{w}}  \right\rVert_2 \nonumber
\end{aligned}
\end{equation}

And over all possible choice of samples $s$ in the batch, we get that 
\begin{align}
    \lVert \nabla L_{B'}(\mathbf{w}) - \nabla L_{B''}(\mathbf{w}) \rVert_2  &\leq \frac{2}{|B|} \max_{i, s} \left\lVert   \frac{\partial z_i(s)}{\partial \mathbf{w}} \right\rVert_2 \nonumber\\
    &= \Delta S_B(\mathbf{w}) \label{eq:sensitivity},
\end{align}
where $z_i(s)$ is the $i$th coordinate of the vector $\mathbf{z}$ in the sample $s$. 

Now, in Step 4 of our protocol the partial derivative vector $\frac{\partial z_i(s)}{\partial \mathbf{w}}$ from Eq.~\ref{eq:app:two-terms} is encoded as:
\[
\left\lfloor r \frac{\partial z_i(s)}{\partial \mathbf{w}} \right\rfloor.
\]
Plugging this instead of the partial derivative  vector in Eq.~\ref{eq:app:two-terms}, we see that the expression remains the same except that now the partial derivative vector has been changed. Therefore, from Eq.~\ref{eq:sensitivity}, we get:
\[
    \lVert \nabla \widetilde{L}_{B'}(\mathbf{w}) - \nabla \widetilde{L}_{B''}(\mathbf{w}) \rVert_2  \leq \frac{2}{|B|} \max_{i, s} \left\lVert   \left\lfloor r \frac{\partial z_i(s)}{\partial \mathbf{w}} \right\rfloor \right\rVert_2,
\]
where $\widetilde{L}$ represents the loss function using the encoded partial derivative vector. Next we use the following proposition:

\begin{proposition}
\label{prop:floor-vec-one}
Let $\mathbf{x} \in \mathbb{R}^m$ and $r$ be a positive integer. Then
\[
\lVert \lfloor r\mathbf{x} \rfloor  \rVert_2 \leq r \lVert \mathbf{x}   \rVert_2 .
\]
\end{proposition}
\begin{proof}
We have
\begin{align*}
   \lVert \lfloor r\mathbf{x} \rfloor  \rVert_2 &= \sqrt{ \sum_{i = 1}^m ( \lfloor r x_i \rfloor )^2} \\
   &\leq \sqrt{ \sum_{i = 1}^m ( r x_i)^2} = r \sqrt{ \sum_{i = 1}^m ( x_i)^2} = r \lVert \mathbf{x}   \rVert_2.
\end{align*}
\end{proof}
Let us denote the vector $\frac{\partial z_i(s)}{\partial \mathbf{w}}$ which maximizes $\left\lVert   \left\lfloor r \frac{\partial z_i(s)}{\partial \mathbf{w}} \right\rfloor \right\rVert_2$ as $\mathbf{z}$. Then, in light of the above proposition and Eq.~\ref{eq:sensitivity}:
\begin{align}
     \lVert \nabla \widetilde{L}_{B'}(\mathbf{w}) - \nabla \widetilde{L}_{B''}(\mathbf{w}) \rVert_2  &\leq \frac{2}{|B|} \max_{i, s} \left\lVert   \left\lfloor r \frac{\partial z_i(s)}{\partial \mathbf{w}} \right\rfloor \right\rVert_2 \nonumber\\
     &= \frac{2}{|B|} \left\lVert   \left\lfloor r \mathbf{z} \right\rfloor \right\rVert_2 \nonumber \\
     &\leq \frac{2}{|B|}r \left\lVert   \mathbf{z} \right\rVert_2 \nonumber \\
     &\leq \frac{2}{|B|}r \max_{i, s} \left\lVert    \frac{\partial z_i(s)}{\partial \mathbf{w}} \right\rVert_2 \nonumber \\
     &= r\Delta S_B(\mathbf{w})\label{eq:true-sensitivity}
\end{align}

\section{Proving Security}
\label{app:security-proof}
We will prove security in the real-ideal paradigm~\cite[\S 23.5]{boneh2020graduate},\cite{canetti2001universal-composable}. Under this framework, we need to consider how we can define ideal functionality for $P_2$. More specifically, $P_2$ applies differentially private (DP) noise at places in the protocol. One way around this is to assume that the ideal functionality applies DP noise to the labels of $P_2$'s dataset at the start, and uses these noisy labels to train the dataset. However, this may cause issues with the amount of DP noise added in the ideal world vs the real world. To get around this, we assume that the ideal functionality does the same as what happens in the real-world, i.e., in each batch, the ideal functionality adds noise according to the sensitivity of the batch. We can then argue that the random variables representing the output in both settings will be similarly distributed. We assume $R$, the number of weights in the last layer, $|B|$, the batch size, and the number of epochs to be publicly known.

\descr{The ideal world.} In the ideal setting, the simulator $\mathcal{S}$ replaces the real-world adversary $\mathcal{B}$. The ideal functionality $\mathcal{F}$ for our problem is defined as follows. The environment $\mathcal{Z}$ hands the following inputs to the ideal functionality $\mathcal{F}$:
\begin{itemize}
    \item \textit{$P_1$'s inputs:} dataset $D_1$, hold-out set $D_\text{hold}$, learning algorithm $\mathcal{A}$, $L_\text{hold}({M_1})$, where $M_1 \leftarrow \mathcal{A}(D_1)$, and the features of the dataset $D_2$, i.e., without the labels, which is the input to party $P_1$.
    \item \textit{$P_2$'s input:} dataset $D_2$.
    \item \textit{Leaked parameters of $M_1$:} the parameters $R$ (number of weights in the last layer), batch size $|B|$, and the number of epochs $n$. 
    \item \textit{System parameters:} the privacy parameter $\epsilon$, and the list of $t$ allowable sensitivities to the ideal functionality $\mathcal{F}$.
\end{itemize}

The ideal functionality $\mathcal{F}$ then proceeds as follows:

\begin{itemize}
    \item $\mathcal{F}$ first sets $D \leftarrow D_1 \cup D_2$.
    \item For each epoch, it samples a random batch $B$ of size $|B|$ from $D$. If $B$ does not contain any sample from $D_2$, it proceeds with backpropagation, after which it moves to the next epoch. Else it proceeds as follows.
    \item It computes the sensitivity of the gradients of the loss function for the current batch.
    \item It then looks up the smallest value $s$ in the list of allowable sensitivity values which is greater than or equal to the computed sensitivity, and generates DP-noise $rs \mathcal{N}(0, \sigma^2 \mathbf{1}_R)$, and adds it to the quantity $N_B(\mathbf{w})$ as $ N_B(\mathbf{w})  + \lfloor rs \mathcal{N}(0, \sigma^2 \mathbf{1}_R) \rfloor = \widetilde{N}_B(\mathbf{w})$.
    \item $\mathcal{F}$ then sends the current batch $B$ and the quantity $\widetilde{N}_B(\mathbf{w})$ to $P_1$. Note that this information is leaked in our protocol to $P_1$.
    \item $\mathcal{F}$ then decodes the quantity as $\widetilde{N}_B(\mathbf{w})/r$ and continues with backpropogation. This is exactly what is done by $P_1$.
    \item If this is the last epoch, the functionality outputs 1 if $L_\text{hold}(M_2) < L_\text{hold}(M_1)$, where $M_2$ is the resulting training model on dataset $D$ using algorithm $\mathcal{A}$.
\end{itemize}

Note that each party's input is forwarded directly to the ideal functionality $\mathcal{F}$ (as described above). Thus the simulator $\mathcal{S}$ may ask the ideal functionality for the inputs to and outputs from a corrupt party. Since we assume the honest-but-curious model, $\mathcal{S}$ cannot modify these values. Further note that the ideal functionality is only leaking the noisy gradients $\widetilde{N}_B(\mathbf{w})$ per batch $B$ to $P_1$, and nothing to $P_2$ (except for the output and the leaked model parameters as input). In our proof, we shall show that the real world protocol's output is statistically indistinguishable from this.  

\descr{The real world.} In the real world, the environment $\mathcal{Z}$ supplies the inputs (both private and public) to and receives the outputs from both parties. Any corrupt party immediately reports any message it receives or any random coins it generates to the environment $\mathcal{Z}$. 

\descr{Simulation when $P_1$ is corrupt.} $\mathcal{S}$ first obtains the inputs to $P_1$ directly from the ideal functionality. 

\begin{itemize}
    \item At some point $\mathcal{Z}$ will generate a control message which results in the real-world $P_1$ receiving the (label) encrypted dataset $D_2$ from real-world $P_2$ (Step 1). To simulate this, $\mathcal{S}$ generates $|D_2|$ fresh samples from the distribution $\mathcal{D}_0$ from Definition~\ref{def:tlwe-assumption}, and adds one to each row of $D_2$ as the purported encrypted label. $\mathcal{S}$ reports this encrypted dataset to $\mathcal{Z}$, just like real-world $P_1$ would do. After this step the two encrypted databases (real and simulated) are statistically indistinguishable due to the TLWE assumption over the torus (Definition~\ref{def:tlwe-assumption}).  
    
    \item At a later point $\mathcal{Z}$ generates a control message resulting in $P_1$ receiving $t$ encrypted $R$-element noise vectors (one for each allowable value of sensitivity) from $P_2$ (Step 6). To simulate this, $\mathcal{S}$ again generates $tR$ fresh samples from $\mathcal{D}_0$, and reports the $t$ encrypted $R$-element vectors as the supposed noise vectors to $\mathcal{Z}$. Notice that, the length of each vector does not depend on the number of elements in the batch belonging to $D_2$, as long as there is at least one. If none are from $D_2$, then $P_1$ will not ask $P_2$ to send a noise vector. Once again, at this step the real and simulated outputs are statistically indistinguishable under the TLWE assumption over the torus.
    \item Lastly, $\mathcal{Z}$ generates a control message resulting in $P_1$ obtaining the blinded and noise-added $R$-element vector $N_B(\mathbf{w})$ (Step 9). $\mathcal{S}$ queries $\mathcal{F}$ to obtain the batch $B$ and the quantity $\widetilde{N}_B(\mathbf{w})$. Note that these are sent to $P_1$ and hence the simulator can ask for these inputs. If the batch $B$ does not contain any sample from $D_2$, $\mathcal{S}$ does nothing. Otherwise, 
    it generates a blind vector $\boldsymbol{\mu}$ (Step 8 of the protocol), by generating fresh coins to generate $R$ elements uniformly at random from $\mathcal{P}$ (the plaintext space). $\mathcal{S}$ reports these coins to $\mathcal{Z}$. $\mathcal{S}$ reports $\widetilde{N}_B(\mathbf{w}) + \boldsymbol{\mu}$ to $\mathcal{Z}$. At this step the real and simulated outputs are \emph{perfectly} indistinguishable due to the fact that the blinds are generated uniformly at random. 
    \item If this is the last epoch, $\mathcal{S}$ sends the output returned by $\mathcal{F}$ to $P_1$, dutifully to $\mathcal{Z}$.
\end{itemize}
Thus, overall the output of the environment is statistically indistinguishable from the ideal case under the TLWE assumption over the torus.

\descr{Simulation when $P_2$ is corrupt.}  The simulation from $\mathcal{S}$ is as follows.
\begin{itemize}
    \item When $\mathcal{Z}$ sends the initial input to $P_2$, $\mathcal{S}$ queries $\mathcal{F}$ to obtain this input. At the same time, $\mathcal{S}$ generates fresh coins and uses them to generate the key $\mathsf{k}$ for the TLWE scheme. $\mathcal{S}$ reports these coins to $\mathcal{Z}$.
    \item For generation of Gaussian noise, $\mathcal{S}$ generates an $R$-element noise vector using fresh coins as $\mathcal{N}(0, \sigma^2 \mathbf{1}_R)$. Then for each $s$ in the list of allowable sensitivities, $\mathcal{S}$ computes $rs\mathcal{N}(0, \sigma^2 \mathbf{1}_R)$. $\mathcal{S}$ then encrypts each of these $t$ vectors under $\mathsf{k}$ as $\left\llbracket \lfloor rs\mathcal{N}(0, \sigma^2 \mathbf{1}_R)  \rfloor \right\rrbracket_{\mathsf{k}}$. $\mathcal{S}$ reports these coins to the environment $\mathcal{Z}$ (as the real-world $P_2$ would do).
    \item At some point $\mathcal{Z}$ generates a control message resulting in $P_2$ receiving the encrypted, blinded and noise-added $R$-element vector $N_B(\mathbf{w})$ (Step 8). $\mathcal{S}$ generates $R$ elements uniformly at random from $\mathcal{P}$, and then encrypts the resulting $R$-element vector under $\mathsf{k}$. $\mathcal{S}$ hands this to $\mathcal{Z}$ as the purported received vector.
    \item If this is the last epoch, $\mathcal{S}$ sends the output returned by $\mathcal{F}$ to $P_2$ dutifully to $\mathcal{Z}$.
\end{itemize}

In this case, in each step, the simulation is perfect. We are using the fact that the blinds used completely hide the underlying plaintext. 

\section{Ratio of the Number of Samples vs Ratio of Model Accuracy}
\label{app:ratio}
\begin{figure}[th]
    \centering
    \includegraphics[width=0.9\linewidth]{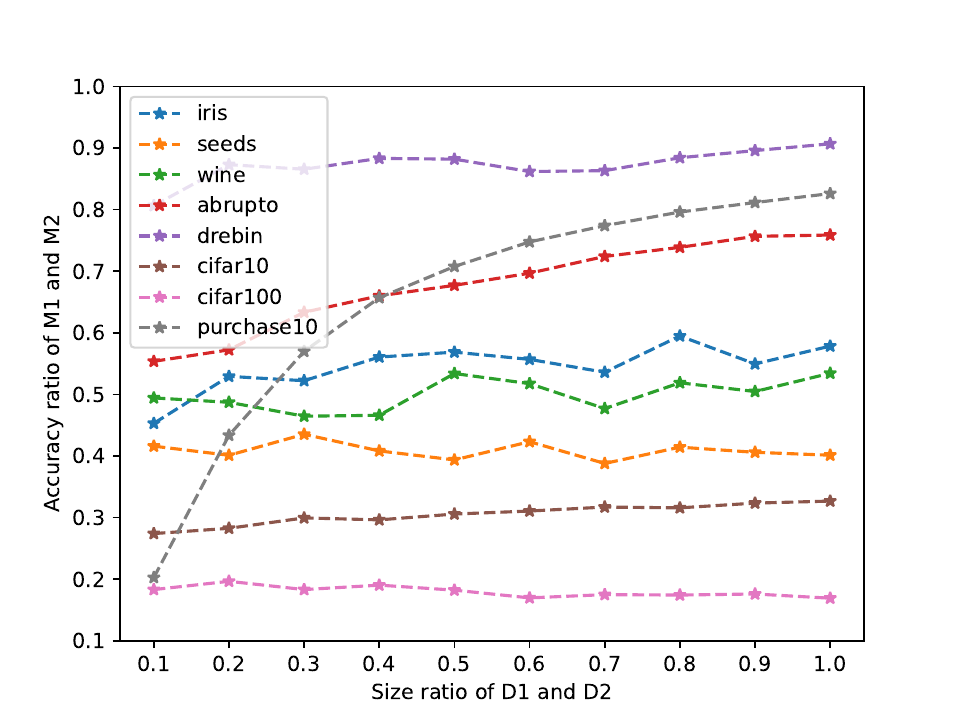}
    \caption{\revise{Ratio of Dataset Size vs Ratio of Model Accuracy.}}
    \label{fig:ratio}
\end{figure}

\revise{Figure~\ref{fig:ratio} shows the effect of the size difference between $D_1$ and $D_2$ on the accuracy of the model $M_2$ over all datasets used in our work. Recall that $M_1$ is trained on $D_1$, and $M_2$ is trained on $D_1 \cup D_2$. The $x$-axis shows the ratio $\frac{\lvert D_{1}\rvert}{\lvert D_{2}\rvert}$ and the $y$-axis shows the ratio $\frac{M_{1} \text{'s Accuracy}}{M_{2} \text{'s Accuracy}}$. For each dataset, we take $D_2$ to be one-third of the entire dataset. The size of $D_1$ is then varied to be $10\%, 20\%, \ldots, 100\%$ of $D_2$ by sampling points from the remaining dataset. $D_\text{hold}$ is constructed to be a balanced dataset and its size remains the same across the different ratios. Each data point is averaged over 100 runs.}  

\revise{We have two key observations. First, the ratio of model accuracy being less than one indicates that $M_{1}$ has worse accuracy than $M_{2}$. Second, the ratio of model accuracy gradually increases and stabilizes as the size of the two becomes similar. This indicates that with the size difference considered in our experiments $M_2$ shows noticeable improvement over $M_{1}$ performance. Hence we conclude that more training data samples result in better training performance. Note that since the deep neural networks are not good on CIFAR datasets~\cite{shokri2017membership,lu2022differentially}, the test accuracy of models on CIFAR-10 and CIFAR-100 could not be enhanced by tuning the dataset size, which is reflected by the two flat curves on the two CIFAR datasets. There are drops in accuracy in some of the dataset as we increase the ratio, more prominently in smaller datasets. This is most probably due to the small number of experimental runs (100 per ratio).}

\section{Neural Network Weights of PyTorch vs Our Implementation}
\label{app:weights-biases}
\descr{Learned Parameters.} Table~\ref{tab:scratch-vs-actual} below shows the weights and biases of the neural network trained via our implementation from scratch versus the PyTorch implementation. 

\begin{table}[!ht]
\centering
\caption{Weights and biases of of the neural network training from scratch versus neural netowrk from PyTorch (hidden neurons: [4,4], batch size: 16, learning rate: 0.1, epochs: 100).}
\label{tab:scratch-vs-actual}
\resizebox{\columnwidth}{!}{
\begin{tabular}{c| c c c c|c}
\toprule
\multicolumn{6}{c}{\bf Input layer to hidden layer 1}\\
\hline
 Type & \multicolumn{4}{c|}{Weights} & Bias \\
\hline
\multirow{4}{*}{Ours} 
   ~&$-0.4916$ & $-$0.6516 & 1.0534 & 0.6013 & $-$0.2006\\
   ~&0.298 & 0.4451 & $-$0.7782 & $-$0.2451 & 0.119\\
  ~&$-$0.4741 & $-$0.7003 & 1.1115 & 0.4583  &$-$0.2123\\
   ~&0.3005 & 0.6029 & $-$0.754 & $-$0.6411 & 0.1407 \\
\hline
\multirow{4}{*}{PyTorch}
   ~&$-$0.4916 & $-$0.6516 & 1.0534 & 0.6013 & $-$0.2006\\
   ~&0.2980 & 0.4451 & $-$0.7782 & $-$0.2451 & 0.119\\
  ~&$-$0.4741 & $-$0.7003 & 1.1115 & 0.4583  &$-$0.2123\\
   ~&0.3005 & 0.6029 & $-$0.754 & $-$0.6411 & 0.1407 \\
\midrule
\multicolumn{6}{c}{\bf Hidden layer 1 to hidden layer 2}\\
\hline
 Type & \multicolumn{4}{c|}{Weights} & Bias \\
 \hline
\multirow{4}{*}{Ours} 
~ & $-$1.4098 & 0.94 &  $-$1.3221 & 1.0272 & 0.4485 \\
~ & 1.1843 & $-$0.5146 & 1.0718 & $-$0.9094 & $-$0.2474 \\
 ~ & $-$0.8463 & 0.5649 &$-$1.0814 & 0.7063 & 0.2926 \\
~ & 1.0468 & $-$0.4571 & 1.0761 & $-$0.6371 & $-$0.1646 \\
\hline
\multirow{4}{*}{PyTorch}
~ & $-$1.4098 & 0.94 &  $-$1.3221 & 1.0272 & 0.4485 \\
~ & 1.1843 & $-$0.5146 & 1.0718 & $-$0.9094 & $-$0.2474 \\
 ~ & $-$0.8463 & 0.5649 &$-$1.0814 & 0.7063 & 0.2926 \\
~ & 1.0468 & $-$0.4571 & 1.0761 & $-$0.6371 & $-$0.1646 \\
\midrule
\multicolumn{6}{c}{\bf Hidden layer 2 to output layer}\\
\hline
 Type & \multicolumn{4}{c|}{Weights} & Bias\\
 \hline
\multirow{4}{*}{Ours} 
~&2.0928 & $-$1.7484 & 1.3045 & $-$1.6554 & -\\
~&$-$0.1631 & 0.3837 & 0.1516 & 0.4605 & - \\
~& $-$2.1381 & 1.5229 & $-$1.2348 & 1.0557 & -\\
\hline
\multirow{4}{*}{PyTorch}
~&2.0928 & $-$1.7484 & 1.3045 & $-$1.6554 & -\\
~&$-$0.1631 & 0.3837 & 0.1516 & 0.4605 & -\\
~& $-$2.1381 & 1.5229 & $-$1.2348 & 1.0557 & -\\
\bottomrule
\end{tabular}
}
\end{table}

\begin{figure}[!ht]
    \centering
    \captionsetup{justification=centering}
    \subfloat[Model from Scratch.]{
    \includegraphics[width=0.25\textwidth]{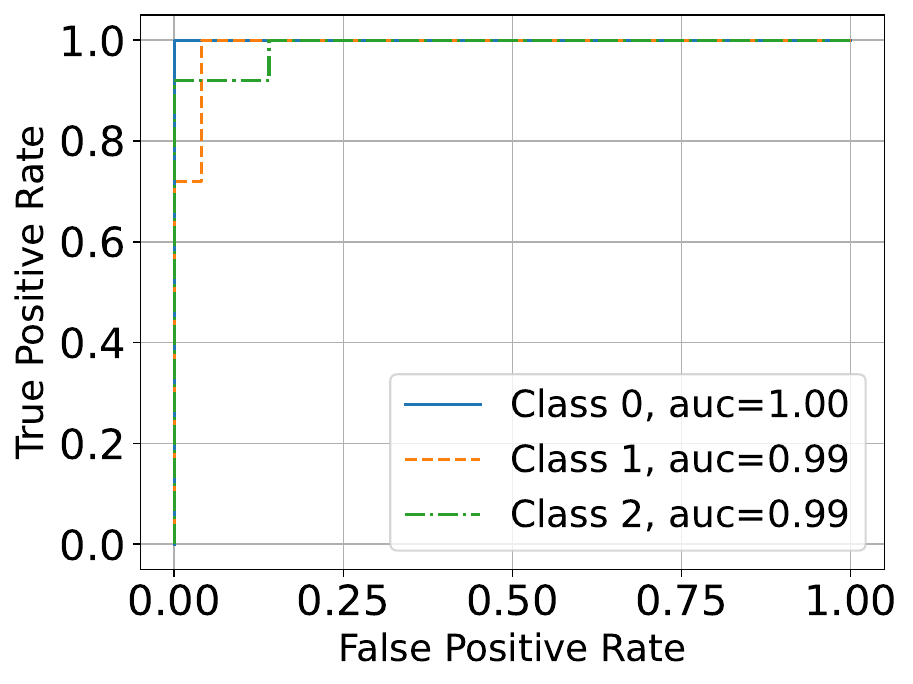}
    }
    \subfloat[PyTorch Model.]{
    \includegraphics[width=0.25\textwidth]{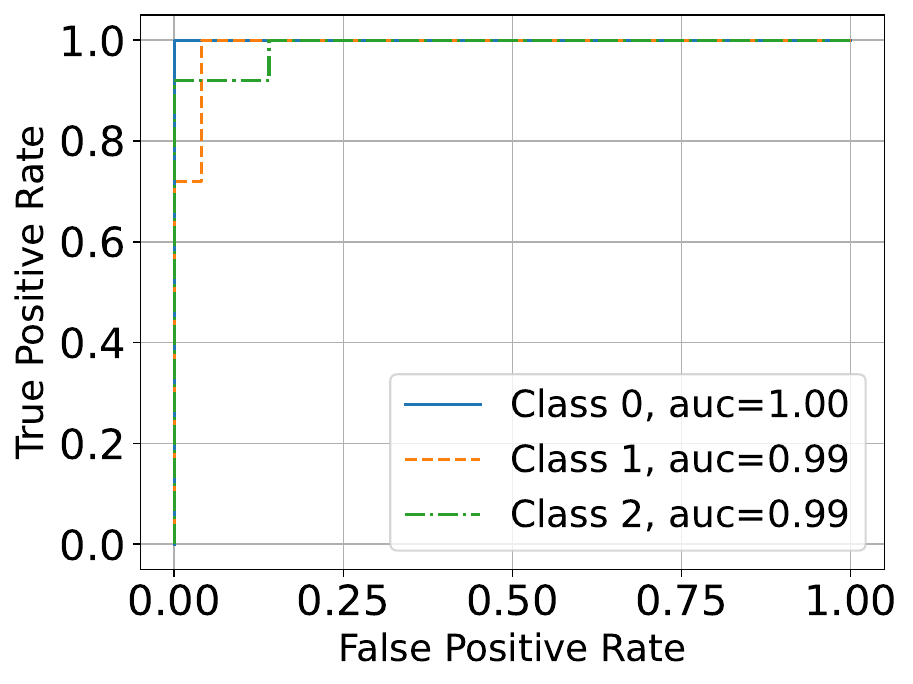}
    \label{subfig:location_priv_leak}
    }
    \caption{ROC Curves of Model from Scratch and PyTorch Model on The Iris Dataset.}
    \label{fig:ROC_from_scratch}
\end{figure}

\descr{Accuracy.} Figure~\ref{fig:ROC_from_scratch} depicts the ROC curve from our model versus the PyTorch implementation on the Iris dataset. As can be seen, our model faithfully reproduces the results from PyTorch. We are therefore convinced that our implementation from scratch is an accurate representation of the model from PyTorch. 

\section{Neural Network Weights of Our Model in Plaintext vs Ciphertext}
\label{app:weights-plain-cipher}
Table~\ref{tab:scratch-vs-encrypted} shows the weights and biases of our neural network model when no encryption is involved (plaintext) versus those of the model implemented through Zama's \concrete{}, i.e., with homomorphic opertions. 

\begin{table}[!ht]
\centering
\caption{Weights and biases of models trained on the Iris dataset in plaintext and ciphertext (hidden neurons: 4, batch size: 16, learning rate: 0.1, epochs: 50).}
\label{tab:scratch-vs-encrypted}
\resizebox{\columnwidth}{!}{
\begin{tabular}{c| c c c c|c}
\toprule
\multicolumn{6}{c}{\bf Input layer to hidden layer 1}\\
\hline
 Type & \multicolumn{4}{c|}{Weight} & Bias \\
 \hline
\multirow{4}{*}{\thead{Model Trained \\ on Plaintext}} 
   ~&0.6511 & 0.0704 & 0.2909  &1.0547 &$-0.3171$ \\
   ~&0.4029 &$-0.4721$ & 1.2437 & 0.6597 &0.472 \\
  ~&$-0.374$ &  0.5  &  $-0.9141$ &$-0.512$& 0.0603\\
   ~&$-0.3569$ & 0.3907 &$-0.9173$ &$-0.8551$& $-0.0134$\\
\hline
\multirow{4}{*}{\thead{Model Trained \\ on Ciphertext}}
    ~&0.6511 & 0.0704 & 0.2909  &1.0546 &$-0.3171$ \\
   ~&0.4029 &$-0.4721$ & 1.2437 & 0.6597 &0.472 \\
  ~&$-0.374$ &  0.4999  &  $-0.9141$ &$-0.5119$& 0.0603\\
   ~&$-0.3569$ & 0.3907 &$-0.9173$ &$-0.8551$& $-0.0135$\\
\midrule
\multicolumn{6}{c}{\bf Hidden layer 1 to output layer}\\
\hline
 Type & \multicolumn{4}{c|}{Weight} & Bias\\
 \hline
\multirow{3}{*}{\thead{Model Trained \\ on Plaintext}} 
~&$-1.0483$ &$-1.8553$ & 1.194 &  1.3057& -\\
~&$-0.4617$ & 0.8366&  0.2772 &$-0.0569$& - \\
~& 0.9567 & 0.822 & $-0.9533$ &$-1.4343$& -\\
\hline
\multirow{3}{*}{\thead{Model Trained \\ on Ciphertext}}
~&$-1.0483$ &$-1.8553$ & 1.194 &  1.3057& -\\
~&$-0.4617$ & 0.8366&  0.2772 &$-0.0569$& - \\
~& 0.9566 & 0.822 & $-0.9533$ &$-1.4343$& -\\
\bottomrule
\end{tabular}
}
\end{table}

\section{Effect of the Number of Classes and Dataset Size on Runtime}
\label{app:runtime:cifar}
{
Table~\ref{tab:final_results4} gives the results on the same features as Table~\ref{tab:final_results3} but this time on the three larger datasets CIFAR-10, CIFAR-100 and Purchase-10. Unfortunately, due to the large total time required to train the entire datasets, these times are extrapolated from the time taken in one epoch. The neural network for these datasets is trained over 70 epochs as the overall accuracy, especially in case of CIFAR-10 and CIFAR-100 is always low (regardless of encryption and differential privacy). The accuracy is determined by evaluating the accuracy when only differentially private noise is added, i.e., with no encryption, since in our earlier experiments in Table~\ref{tab:final_results3} the accuracy with and without encryption match. We acknowledge that the times mentioned in the table are prohibitive. However, once again, this is largely due to the slower time taken by our implementation over the PyTorch implementation. On CIFAR-10, CIFAR-100 and Purchase-10, our implementation is approximately 154, 185, and 71 times slower than that of PyTorch. This means that if we were able to implement our protocol on PyTorch, even without the other speedups mentioned in Section~\ref{sub:limitations}, the protocols would take just over 4 hours for CIFAR-10, one day and 14 hours for CIFAR-100, and one day and 2 hours for Purchase-10. These numbers are substantially better than an end-to-end FHE solution. We also note that the pre-processing time is very large for these datasets. Thus, the time and communication complexity can further be improved by a different noise adding mechanism which allows multiplication of sensitivity with unit noise without breaking the differential privacy guarantee. Lastly, we note that the runtime increases with both the size of the dataset and the number of classes. This is because the homomorphic operations are linear in the number of classes $K$, as mentioned in steps (3) and (4) of the protocol.
}

\begin{table*}[ht]
\centering
\caption{\revise{Extrapolated training time and accuracy of $M_1$ (dataset $D_1$), $M_2$ (dataset $D_1 \cup D_2$) in the clear, $\widetilde{M}_2$ (dataset $D_1 \cup D_2$) through our protocol, $M^{\text{Py}}_2$ (dataset $D_1 \cup D_2$) implemented in PyTorch in the clear, and ${M}^{\text{RR}}_2$ (dataset $D_1 \cup D_2$) with randomized response (hidden neurons: 128, batch size: 32, learning rate: 0.01, epochs: 70, weight decay: 0.001).}}
\label{tab:final_results4}
\resizebox{\textwidth}{!}{
\begin{tabular}{c|c| c c| c c | c | c c |c c | c}
\toprule
\multirow{3}{*}{Dataset} & \multirow{2}{*}{$\epsilon$}  &  \multicolumn{2}{c|}{\thead{Model $M_1$ \\ Plaintext, no DP}} & \multicolumn{2}{c|}{\thead{Model $M_2$\\ Plaintext, no DP}} &  \multicolumn{3}{c|}{\thead{Model $\widetilde{M}_2$\\ Our protocol}} & \multicolumn{2}{c}{\thead{Model $M^{\text{Py}}_2$\\ PyTorch baseline}} &\multicolumn{1}{|c}{\thead{Model $M^{\text{RR}}_2$ \\ Randomized Response}}  \\
\cline{3-12}
 & & Time (s) & Test Acc. & Time (s) & Test Acc. & \thead{ $t$-list time (s) \\ per epoch \& total} &Gap Time (s) & Test Acc. & Time (s) & Test Acc.  & Test Acc.\\ 
\hline
\multirow{4}{*}{CIFAR-10} 
& 1 & \multirow{4}{*}{1m48s}&\multirow{4}{*}{0.3032}&\multirow{4}{*}{1h59m27s} & \multirow{4}{*}{0.4752} & \multirow{4}{*}{\shortstack[c]{6d12h37m43s\\ $\approx$ 456d20h}}&26d16h & 0.1709 & \multirow{4}{*}{46.5314} & \multirow{4}{*}{0.4658} & 0.1967\\ 
&10 & &&& &&26d16h&0.2281&& & 0.4610\\ 
&\underline{100}& &&& &&26d16h&0.3168&& & 0.4712\\ 
&\underline{1000}& &&& &&26d16h&0.3544&& & 0.4721\\ \hline

\multirow{4}{*}{CIFAR-100} 
&1 &\multirow{4}{*}{1m46s}&\multirow{4}{*}{0.0572} &\multirow{4}{*}{2h02m18s}&\multirow{4}{*}{0.2082}&\multirow{4}{*}{\shortstack[c]{69d02h\\ $\approx$ 4836d01h}} &294d16h&0.0146&\multirow{4}{*}{39.5251}&\multirow{4}{*}{0.2084}& 0.0139\\
&10& &&& &&294d16h&0.0334&& & 0.2061\\ 
&\underline{100}& &&& &&294d16h&0.0950&& &0.2084 \\ 
&\underline{1000}& &&& &&294d16h&0.1088&& & 0.2084\\  \hline

\multirow{5}{*}{Purchase-10} 
&1 &\multirow{5}{*}{1m46s}&\multirow{5}{*}{0.911} &\multirow{5}{*}{2h03m32s}&\multirow{5}{*}{0.9742}&\multirow{5}{*}{\shortstack[c]{21d13h\\ $\approx$ 1510d3h}} &79d20h&0.5970&\multirow{5}{*}{01m44s}&\multirow{5}{*}{0.9742} & 0.8654\\ 
&10& &&& &&79d20h&0.8025&&& 0.9684 \\ 
&100& &&& &&79d20h&0.8912&&&0.9714  \\ 
&\underline{125}& &&& &&79d20h&0.9312&&&0.9710  \\ 
&\underline{1000}& &&& &&79d20h&0.9652&&&0.9721  \\ 

\bottomrule
\end{tabular}
}

\end{table*}

\end{document}